\documentclass[lettersize,journal]{IEEEtran}
\usepackage{amsmath,amsfonts}
\usepackage{booktabs}
\usepackage{algorithm2e}
\usepackage{algpseudocode}
\usepackage{array}
\usepackage[caption=false,font=normalsize,labelfont=sf,textfont=sf]{subfig}
\usepackage{textcomp}
\usepackage{stfloats}
\usepackage{url}
\usepackage{verbatim}
\usepackage{graphicx}
\usepackage{cite}
\usepackage{caption}
\usepackage{multirow}
\usepackage{tabularx}
\usepackage{amssymb}
\usepackage{makecell}
\RestyleAlgo{ruled}

\newcommand{\mcl}[1]{\mathcal{#1}}

\begin{document}

\title{Low-frequency Image Deep Steganography: Manipulate the Frequency Distribution to Hide Secrets with Tenacious Robustness}

\author{Huajie~Chen,
        Tianqing Zhu*,
        Yuan Zhao,
        Bo Liu,
        Xin Yu,
        and Wanlei Zhou
\thanks{*Tianqing Zhu is the corresponding author.}
\thanks{Huajie Chen, Tianqing Zhu, Yuan Zhao, and Bo Liu are with the Centre for Cyber Security and Privacy and the School of Computer Science, University of Technology Sydney, Sydney, NSW 2007, Australia (e-mail: huajie.chen@student.uts.edu.au; tianqing.zhu@uts.edu.au; yuan.zhao@student.uts.edu.au; bo.liu@uts.edu.au;).}
\thanks{Xin Yu is with the School of Information Technology and Electrical Engineering
, University of Queensland, Brisbane, QLD 4067, Australia (email: xin.yu@uts.edu.au)}
\thanks{Wanlei Zhou is with the Institute of Data Science, City University of Macau, Macao SAR, China (e-mail: wlzhou@cityu.edu.mo).}}



\maketitle

\begin{abstract}
Image deep steganography (IDS) is a technique that utilizes deep learning to embed a secret image invisibly into a cover image to generate a container image. However, the container images generated by convolutional neural networks (CNNs) are vulnerable to attacks that distort their high-frequency components. To address this problem, we propose a novel method called Low-frequency Image Deep Steganography (LIDS) that allows frequency distribution manipulation in the embedding process. LIDS extracts a feature map from the secret image and adds it to the cover image to yield the container image. The container image is not directly output by the CNNs, and thus, it does not contain high-frequency artifacts. The extracted feature map is regulated by a frequency loss to ensure that its frequency distribution mainly concentrates on the low-frequency domain. To further enhance robustness, an attack layer is inserted to damage the container image. The retrieval network then retrieves a recovered secret image from a damaged container image. Our experiments demonstrate that LIDS outperforms state-of-the-art methods in terms of robustness, while maintaining high fidelity and specificity. By avoiding high-frequency artifacts and manipulating the frequency distribution of the embedded feature map, LIDS achieves improved robustness against attacks that distort the high-frequency components of container images.
\end{abstract}


\section{Introduction}
\label{sec:intro}

Steganography \cite{chanu2012image} is a technique for hiding a message within another message, such as text, audio, or images, in an imperceptible manner. Image steganography, as a subclass of steganography, aims to hide a secret image in a cover image without being noticed, resulting in a container image. The secret image can then be extracted from the container image when needed.
Traditional image steganography algorithms \cite{jung2015steganographic, rawat2013steganography, bhardwaj2016image} achieve this goal by modifying the least significant bits of the cover image. Some other algorithms \cite{baby2015novel, tsui2008color} use transform techniques, such as discrete wavelet transform (DWT) and multidimensional Fourier transformation, to hide the secret image. Although these methods achieve good imperceptibility, their capacity is limited, and they are not sufficiently robust against transmission loss and malicious attacks.

\begin{figure}[t!]
    \centering
    \includegraphics[width=.49\textwidth]{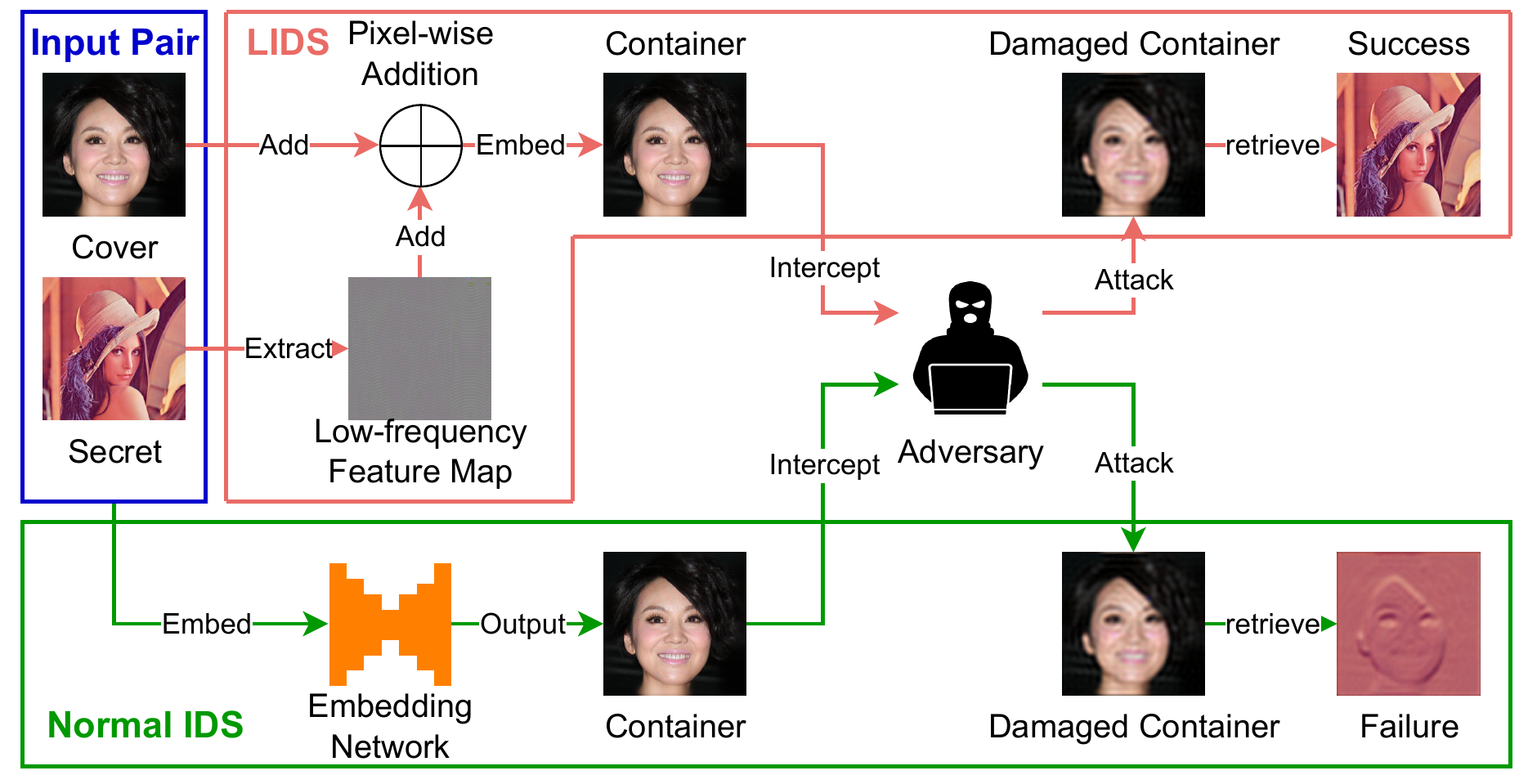}
    \caption{
        Main Idea.
        LIDS extracts a low-frequency feature map from the secret image, and add it to the cover image to generate the container image.
        This avoids the high-frequency artifacts of the CNNs.
        After being attacked, the damaged container image can still be used to recover the secret image due to the enhanced robustness.
        }
    \label{fig:main_idea}
\end{figure}

In recent years, image steganography has evolved into the next generation known as image deep steganography (IDS), which achieves significantly improved capacity and fidelity based on deep learning techniques. IDS was first proposed by Baluja \cite{baluja2017hiding}, who utilized a deep learning model to perform image steganography. Since then, numerous works have been developed to improve the functionalities of IDS.
IDS can generally be categorized into two subclasses based on their network structure. One subclass \cite{zhang2020udh, weng2019high} uses a separated embedding network and retrieval network, while the other subclass \cite{jing2021hinet, lu2021large} employs a unified revertible network. Both types of methods are capable of concealing at least one secret image within a cover image of the same size covertly. However, the robustness of such methods is still insufficient due to the intrinsic properties of convolutional neural networks (CNNs).

Recent studies have shown that images synthesized by convolutional neural networks (CNNs) tend to have an abnormal distribution in their high-frequency domain \cite{durall2020watch, liu2022making}, and a theoretical explanation for this phenomenon has been proposed in \cite{khayatkhoei2022spatial}. In the current mainstream IDS methods \cite{lu2021large, jing2021hinet, weng2019high}, the secret image is intentionally embedded into the high-frequency components of the container image to achieve high fidelity. This is because human eyes are less sensitive to changes in the high-frequency components of images.

However, this strategy also makes the hidden secret image vulnerable to attacks that jeopardize the high-frequency components of the container image, such as JPEG compression and low-pass filtering. These attacks can easily remove the hidden secret image while maintaining a negligible cost of perceptual quality. Previous works \cite{xiang2008invariant, daren2001dwt} have proposed invisibly embedding watermarks into the low-frequency domain of target images to enhance robustness. This could be a solution to address the insufficient robustness of current IDS methods, but it is challenging to implement due to the inherent properties of CNNs. Changes in the high-frequency domain of images are difficult to detect by human eyes, whereas perturbations in the low-frequency domain are more visible. Therefore, if an adversary aims to remove the hidden secret image embedded in the low-frequency domain, the quality cost of the images would be non-negligible. While the current mainstream IDS methods achieve high fidelity by embedding the secret image into the high-frequency components of the container image, this approach also makes the hidden secret image vulnerable to certain attacks. Invisibly embedding the secret image into the low-frequency domain could enhance robustness, but it presents implementation challenges. Further research is needed to develop new approaches that balance fidelity and robustness in IDS methods.

\textbf{Our Work.}
In pursuit of the aforementioned goals, we propose an approach to overcome the aforementioned limitations. Our approach involves the following steps: Firstly, the high-frequency artifacts introduced by the network must be circumvented. Secondly, the secret image is incorporated into the low-frequency components of the container image to enhance its robustness, akin to previous low-frequency image watermarking methods. Finally, the frequency range should be adjustable to enable the creator of the container image to balance the trade-off between robustness and fidelity. By carefully selecting the frequency range, the creator can minimize perceptual costs while simultaneously maximizing the container image's robustness against attacks, and vice versa.

Based on the motivation, we propose a method named Low-frequency Image Deep Steganography, abbreviated as LIDS.
The main idea of LIDS is depicted in Fig. \ref{fig:main_idea}.
Specifically, we employ an embedding network and a retrieval network as the two major building blocks in LIDS.
Different from other IDS methods, this embedding network does not directly output the container image so as to get rid of the high-frequency artifacts brought by the CNNs.
The embedding network firstly extracts a feature map from the secret image, and add it to the cover image to generate the container image.
The feature map is regulated by a frequency loss function, such that the feature map's frequency distribution mainly concentrates on the low-frequency domain.
To further boost the robustness, we insert an attack layer between the embedding network and the retrieval network that randomly launches attacks on the container image.
The retrieval network can thus learn to retrieve the recovered secret image with high quality from the damaged container image.
Additionally, to make the retrieval network differentiate the container image from the clean image, we design a clean loss term.
This prevents the retrieval network from outputting the secret image if it receives the clean image.
This ability is defined as specificity in this context.

Our main contributions are listed as follows:
\begin{itemize}
    \item We propose LIDS that embeds the secret image into the low-frequency component of the cover image. 

    \item We design a new deep learning framework that allows the embedding network to manipulate the frequency distribution of its extracted feature map.

    \item Our experiments demonstrate that LIDS has outperformed all the state-of-the-art methods in terms of robustness under extreme attacks, while preserving high fidelity and specificity.
\end{itemize}

The rest of the paper is organized as follows.
In Section \ref{sec:related}, we list some of the relevant studies in the domain of image steganography and other related fields.
Section \ref{sec:threat} introduce the adversary's motivation, background knowledge, and possible attack methods.
The details of the method are illustrated in Section \ref{sec:method}.
How we design and conduct the experiments is documented in Section \ref{sec:exp}, along with the discussions about the results.
The analysis of the method is in Section \ref{sec:analysis}, and lastly, we draw conclusions in Section \ref{sec:con}.

\section{Related Work}
\label{sec:related}

The notations used in the rest of the paper are listed in Tab. \ref{tab:notations}.

\begin{table}[ht!]
    \caption{Notations}
    \label{tab:notations}
    \begin{tabularx}{.48\textwidth}{
    |>{\centering\arraybackslash}m{.05\textwidth}
    |>{\arraybackslash}m{.374\textwidth}
    |}
    \hline
    
    \multicolumn{1}{|c|}{\textbf{Notation}}
    & 
    \multicolumn{1}{c|}{\textbf{Definition}}
    \\\hline
    
    $c$
    & 
    The cover image.
    \\\hline

    $s$
    & 
    The secret image.
    \\\hline

    $c'$
    & 
    The container image where the secret image is embedded, and it is visually similar to $c$.
    \\\hline

    $s'$
    & 
    The recovered secret image derived from $c'$.
    \\\hline

    $E$
    & 
    The embedding network that embeds $s$ in $c$ to yield $c'$.
    \\\hline

    $R$
    & 
    The retrieval network that retrieves $s'$ from $c'$.
    \\\hline

    $q$
    & 
    The feature map extracted from $s$ via $E$ that is then added onto $c$ to yield $c'$.
    \\\hline

    $f$
    & 
    The low-pass filter that filters out the high-frequency components in the input.
    \\\hline

    $q'$
    & 
    The low-frequency feature map processed by $f$.
    \\\hline

    $A$
    & 
    The attack layer that distorts $c'$ to hurdle the secret retrieval.
    \\\hline

    $c'_a$
    & 
    The damaged container image processed by $A$.
    \\\hline

    $s^N$
    & 
    The null image, i.e., a pure black image.
    \\\hline

    $\hat{s}^N$
    & 
    The null image extracted from the input that does not contain the embedded secret.
    \\\hline

    $l_{\text{emb}}$
    & 
    The embedding loss.
    \\\hline

    $l_{\text{freq}}$
    & 
    The frequency loss.
    \\\hline

    $l_{\text{ret}}$
    & 
    The retrieval loss.
    \\\hline

    $l_{\text{cln}}$
    & 
    The clean loss.
    \\\hline

    $\mcl{F}$
    & 
    The Focal frequency loss function.
    \\\hline

    $\mathcal{L}$
    & 
    The total loss.
    \\\hline
    
    \end{tabularx}    
\end{table}

\textbf{Image Steganography.}
Image steganography \cite{cheddad2010digital} aims to embed a secret image in a cover image to yield a container image.
The embedding algorithm takes in the secret image and the cover image, and outputs the container image that carries the secret image.
The container image is almost visually identical to the cover image.
If needed, the secret image can be retrieved from the container image.

We use three terms to evaluate an image steganography method, namely, fidelity, robustness, and specificity.
High fidelity indicates high visual quality of the container image and the recovered secret image.
Robustness means that the secret image ought to be validly retrieved from the container image that suffers from transmission loss and malicious attacks.
Lastly, the secret image must no be retrieved from an image that does not contain the corresponding signal, which is defined as specificity.

Traditional algorithms conceal the secret image with different strategies, including the least significant bits (LSB) \cite{jung2015steganographic, rawat2013steganography, bhardwaj2016image}, pixel value differencing (PVD) \cite{zhang2004vulnerability}, histogram shifting \cite{qin2012inpainting}, discrete Fourier transform (DFT) \cite{alturki2001secure}, discrete wavelet transform (DWT) \cite{baby2015novel, tsui2008color}, etc.
These methods have achieved promising fidelity.
However, they do not support large embedding capacity, and they fail to provide adequate robustness.

\textbf{Image Deep Steganography (IDS).}
Distinct from the traditional algorithms, IDS utilizes deep learning to accomplish the same task with greatly enhanced embedding capacity.
Baluja \cite{baluja2017hiding} proposes the first method to perform image steganography via deep learning, successfully embedding an entire secret image in a cover image sharing the same size.
In this method, a preparation network extracts the features from a secret image before an embedding network incorporates the features into a cover image to produce a container image.
Eventually, a retrieval network retrieves the secret image from the container image.

Consequently, a number of deep steganography techniques have been developed, and they can be roughly divided into two subclasses, where the boundary is drawn by using a unified network or not.
One subclass \cite{zhang2020udh, weng2019high} uses separate embedding and retrieval networks.
The networks usually have different structures and parameters, but they are trained simultaneously.
Moreover, by employing an extra discriminative network, \cite{hayes2017generating, volkhonskiy2020steganographic} have further improved the imperceptibility of the hidden content.
The other subclass \cite{jing2021hinet, lu2021large} adopts a revertible network as their backbone.
This network serves as both the embedding and retrieval network, achieving high fidelity in both embedding and retrieval process.
Some other methods \cite{liu2022image} choose a completely different strategy where the embedding process is replaced by synthesizing a container image based on the secret information using a generative adversarial network.
Although these promising IDS methods have achieved outstanding fidelity and capacity, the robustness of them is somewhat insufficient due to the inherent property shared by the CNN-based networks \cite{durall2020watch, liu2022making, khayatkhoei2022spatial}.

Remarkably, it is very efficacious to insert a well-designed attack layer in the network to boost the network's robustness against transmission loss and malicious attacks.
The experiments in \cite{wengrowski2019light, tancik2020stegastamp} indicate that the attack layer plays a crucial role in terms of the enhancement of robustness.
Additionally, \cite{xu2022robust} inserts a distortion-guided modulation over flow-based blocks to enhance the robustness of the network.
The secret image can be retrieved from damaged container images with considerably high visual quality.

\textbf{Attack on Images.}
The goal of an adversary is to remove the embedded secret image from an intercepted container image so that it can sabotage the image steganography, even at the expense of the container image's quality.
The two main types of image attacks currently in use are geometric deformation attacks and additive noise-like manipulation attacks \cite{hu2020cover}.
JPEG compression, low-pass filtering, Gaussian blurring, etc. are examples of common additive noise-like manipulation attacks.
Geometric deformation attacks like rotation, scaling, resizing, cropping, flipping, etc., are harder to fend off, because even minor geometric alterations to the container image could significantly lower the possibility to recover the hidden content \cite{Alghoniemy2004ImgWm}.

In addition, techniques that use neural networks to simulate attacks and act as attack layers during training have been developed to increase the robustness of the image watermarking techniques.
For instance, Luo et al. \cite{luo2020distortion} trained neural networks that mimic various attacks, such as Gaussian blur and color change, and incorporated them into the training process of the image steganography network to increase its robustness.
In order to simulate the digital-to-physical transformation process and produce reliable adversarial examples, Jan et al. \cite{jan2019connecting} used an image-to-image translation network.

\textbf{Insertion of the Attack Layer.}
Remarkably, it is very efficacious to insert a well-designed attack layer in the network to boost the network's robustness against transmission loss and malicious attacks.
The experiments in \cite{wengrowski2019light, tancik2020stegastamp} indicate that the attack layer plays a crucial role in terms of the enhancement of robustness.
Additionally, \cite{xu2022robust} inserts a distortion-guided modulation over flow-based blocks to enhance the robustness of the network.
The secret image can then be retrieved from damaged container images with considerably high visual quality.

\section{Threat model and problem definition}
\label{sec:threat}

\textbf{Two Parties, Two Motivations.}
In this study, we consider two parties totally, namely, the defender and the adversary.
The defender is the creator of the container image, whose goal is to ensure that the container image withstand any type of distortion and malicious attack, such that the hidden secret can be successfully retrieved.
The adversary, on the contrary, attempts to prevent the valid retrieval of the hidden secret while maintaining the perceptual quality of the container image.
Thus, the adversary uses image attack methods to sabotage the hidden secret, and reduces the image quality cost as possible as s/he can.

\textbf{Adversary's Background Knowledge.}
We assume that the adversary can directly access the container images.
However, the structure and the parameters of the embedding and retrieval networks are kept from the adversary.
Additionally, the adversary does not know the secret images.
Generally, the adversary would perform steganalysis \cite{fridrich2012rich} to detect whether or not there is any hidden secret in a given image.
Thereafter, the adversary decides if s/he launches attacks to purge the hidden secret.
However, we assume that the adversary evenly treats all the given images as container images, i.e., the adversary launches attacks without steganalysis.
Because even if the attacker is aware that the given image is a container image, there is no difference if the adversary cannot prevent the secret retrieval at a low perceptual cost.

\textbf{Adversary's Attacks.} 
Common additive noise-like manipulation attacks includes JPEG compression, low-pass filtering, Gaussian blurring, etc.
These attacks mainly distort the high-frequency components of the image, where the hidden secrets are usually embedded by the mainstream IDS methods.
Hence, these attacks are effective and can maintain the perceptual loss to an extent.
In contrast, geometric deformation attacks such as resize cropping, rotation, flipping, rotation can easily remove the hidden secret, because they drastically modify the image's features.
However, the cost of geometric deformation attacks is non-negligible, which can be observed in the experiment section.

Apart from the aforementioned traditional algorithm-based attacks, the adversary can also use the network-based attacks that mimicking the attack algorithms such as Gaussian blurring and JPEG compression.
Nontheless, we do not consider such types of attacks in this study, because the network-based attack yields the similar results as those of the tradition algorithm-based attacks.

\section{Method}
\label{sec:method}

\begin{figure*}[t!]
    \centering
    \includegraphics[width=\textwidth]{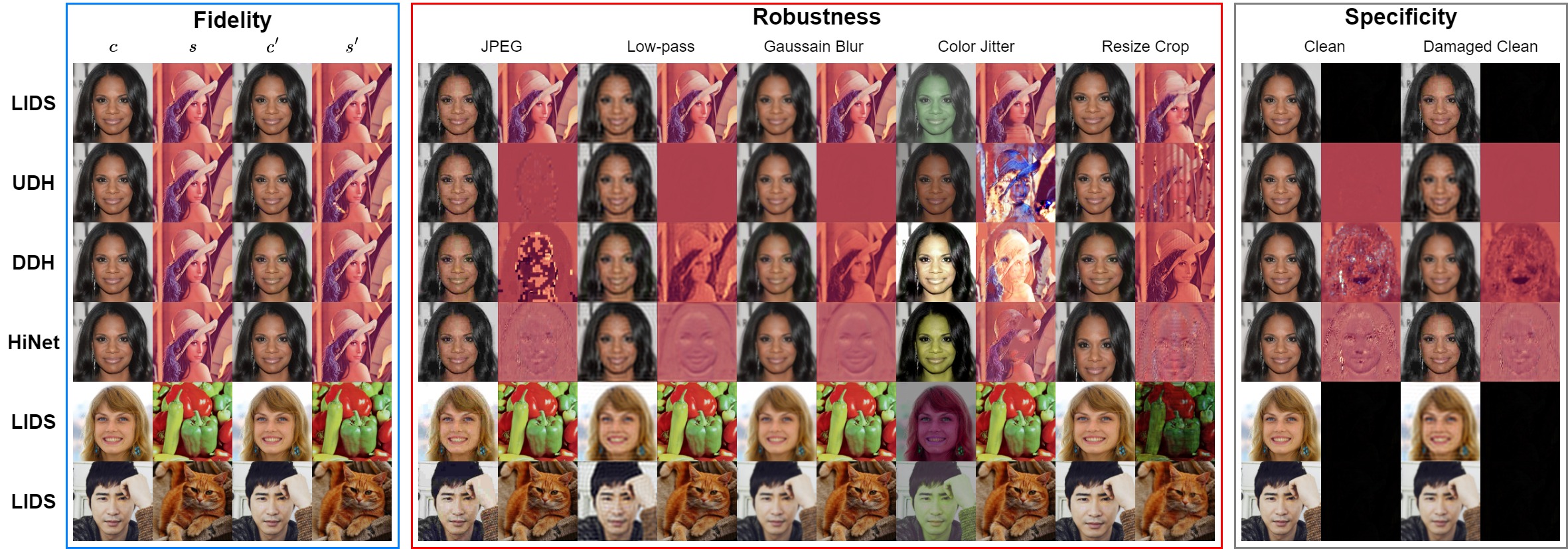}
    \caption{Framework of LIDS.
        During \textbf{embedding}, the embedding network extracts the feature map from the secret image.
        The feature map is directly added to the cover image to yield the container image.
        By adjusting the range of the Fourier Transform spectrum to cut off from the feature map in the low-pass filter, the user can manipulate the frequency distribution of the feature map.
        In the \textbf{retrieval} process, the retrieval network retrieves the recovered secret image from the damaged container image.
        If the retrieval network receives a clean image, it will produce a recovered null image.
    }
    \label{fig:framework}
\end{figure*}

\subsection{Overview}
As depicted in Fig. \ref{fig:framework}, the main objective of LIDS is to robustly embed one secret image into an arbitrary cover image so as to generate a container image.
After being attacked, the container image can still support high-quality retrieval of the secret image.
Additionally, LIDS should be capable of differentiating the container image from the clean image, i.e., an image without hidden content in it.

Briefly, the embedding network $E$ takes in the secret image $s$ to extract a low-frequency feature map $q$.
Then, $q$ is directly added to the cover image $c$ to yield the container image $c'$.
The attack layer launches attacks on $c'$ to yield a damaged container image $c'_a$.
$c'_a$ is then fed into the retrieval network $R$.
A valid recovered secret image $s'$ is eventually retrieved from $c'_a$ by $R$.
When $R$ receives a clean image, $R$ outputs a null image $s^N$ that is simply a pure black image.

\subsection{Attack and Defense}
In general, we categorize the attack into two classes, namely, the additive noise-like manipulation attacks and the geometric deformation attacks.
We take different strategies to face the attacks.

\textbf{Additive Noise-like Manipulation.}
The goal of JPEG compression is to reduce the file size of an image while minimizing the loss of image quality or perceptual fidelity.
The JPEG compression process can be divided into three stages: color space conversion, transform coding, and quantization.
In the second step, discrete cosine transform (DCT) is applied to the image so as to remove the high-frequency components that are less important to human eyes.

An image low-pass filter is a mathematical operation applied to an image to enhance its low-frequency components or smooth out high-frequency noise. Low-frequency components represent the gradual changes in the pixel values and are usually associated with the overall brightness and contrast of the image.

Gaussian blurring is a common image processing technique used to reduce the noise in an image and to smooth out its details. It is achieved by convolving the image with a Gaussian filter kernel, which is a 2D bell-shaped function that represents the probability distribution of a Gaussian random variable.
The Gaussian blurring technique is often used in image processing to remove high-frequency noise such as salt-and-pepper noise and to smooth out the sharp edges and fine details in the image.

Color jittering is a technique used in digital image processing and computer graphics to add variations of color to an image. It involves changing the color values of the pixels in the image by adding a small amount of random noise to them. The amount of jittering applied to each color channel is controlled by a jittering intensity parameter.
Color jittering can be applied to the hue, saturation, and brightness channels of an image separately, or to all three channels simultaneously. The jittering intensity parameter controls the amount of jittering applied to each channel. A low jittering intensity will result in subtle variations of color, while a high jittering intensity can produce more drastic changes.

These four types of attacks are selected as the representatives for the additive noise-like manipulationa attacks, for they can significantly affect the high-frequency component of the image.

\textbf{Geometric Deformation}
Even though there are several types of geometric deformation attacks, we select resize cropping as the representative attack, for it reshape the entire image drastically and is thus the most powerful attack.

Resize cropping is a technique used in image processing to adjust the size and aspect ratio of an image.
Resize cropping combines both resizing and cropping techniques to adjust the size and aspect ratio of an image while minimizing distortion and loss of important details. It involves selecting a rectangular area of the original image that has the desired aspect ratio and resizing it to the desired size. This results in a new image that has the desired size and aspect ratio while preserving the important details in the image.

\textbf{Defenses.}
In order to mitigate additive noise-like manipulation attacks, we utilize a frequency manipulation technique for embedding the secret image. 
This technique is designed to prevent modification of the high-frequency components that may be caused by such attacks.
To address geometric deformation attacks, we insert an attack layer between the container image and the retrieval network. 
This attack layer is responsible for randomly applying not only geometric deformations, but also additive noise-like attacks in order to increase the robustness of the proposed LIDS.

\subsection{Embedding}

\paragraph{Extraction of the Low-frequency Feature Map.}
To ensure that the secret image $s$ is correctly embedded into the low-frequency components of the container image $c$, we must make the embedding network $E$ to extract the feature map $q$ that has a frequency distribution mainly concentrating on the low-frequency domain.

We first use $E$ to extract $q$ from $s$, defined as
\begin{equation}
    \begin{aligned}
        q = E(s).
    \end{aligned}
\end{equation}
$q$ shares the same size as that of $s$.

Then, we apply a low-pass filter $f$ to $q$ to get its counterpart $q'$ that is deprived of high-frequency information, defined as
\begin{equation}
    \begin{aligned}
        q' = f(q, d),
    \end{aligned}
\end{equation}
where $d$ is the range of the Fourier Transform spectrum to cut off from $q$.
A smaller $d$ indicates that less high-frequency information is preserved after the filtering.

In order to make $E$ directly extract $q$ that has a similar distribution as that of $q'$, we use a frequency loss term to regulate $E$.
The frequency loss term is defined as
\begin{equation}
    \begin{aligned}
        l_{\text{freq}} = \mcl F(q, q').
    \end{aligned}
\end{equation}
Here, $\mcl F(\cdot, \cdot)$ denotes the Focal frequency loss function \cite{jiang2021focal}, which is defined as
\begin{equation}
    \begin{aligned}
        \mcl F(q, q') = \frac{1}{WH} 
                       \sum_{u=1}^{W}
                       \sum_{v=1}^{H}
                       w(u, v)
                       |F_q(u, v) - F_{q'}(u, v)|^2,
    \end{aligned}
\end{equation}
where $q$ and $q'$ denote two feature maps of the same size;
$W$ and $H$ are the width and height of $a$ and $b$;
$u$ and $v$ represent the coordinate of a spatial frequency on the frequency spectrum;
$w(u, v)$ serves as the weight for the spatial frequency at $(u, v)$;
$F_{a/b} $ is the spatial frequency value of $a/b$ at $(u, v)$.
This loss term is considered as a weighted average of the frequency distance between $q$ and $q'$, and can thus guide $E$ to extract the desired $q$.

\paragraph{Generation of the Container Image.}
To avoid the problem caused by the intrinsic property of the neural network, we choose not to use the network to directly output $c'$.
Instead, inspired by \cite{zhang2020udh}, we add $q$ directly to $c$ to create $c'$, defined as
\begin{equation}
    \begin{aligned}
        c' = c + q.
    \end{aligned}
\end{equation}

To maintain the visual quality of the container image, we compute the embedding loss that is defined as
\begin{equation}
    \begin{aligned}
        l_{\text{emb}} = \frac{1}{CWH} \sum_i^C \sum_j^W \sum_k^H
                        (c_{ijk} - c'_{ijk})^2,
    \end{aligned}
    \label{eq:l_emb}
\end{equation}
where $C$, $W$ and $H$ respectively denote the number of channels, width, and height of $c$ and $c'$.
This is a loss term that directly measures the pixel-wise distance between $c$ and $c'$, which can help improve the fidelity.

\begin{figure}[t!]
    \centering
    \includegraphics[width=.49\textwidth]{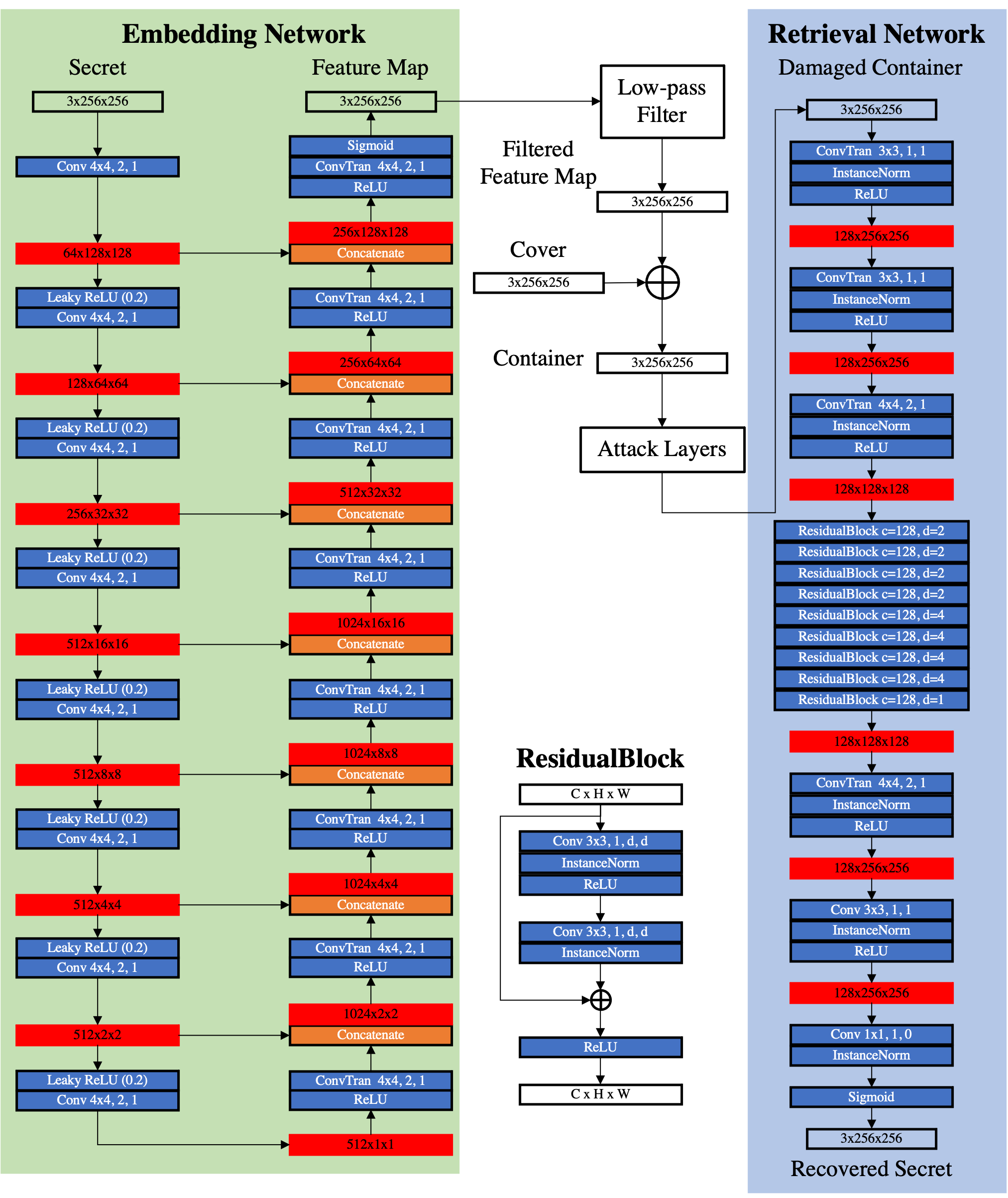}
    \caption{
        Network Structure.
        On the left, the embedding network down-samples the secret image to features of size $512 \times 1 \times 1$, and then, up-samples the features to a feature map.
        The feature map is added to the cover image to get the container image.
        The retrieval network (on the right) down-samples the container image first, and passes it through several residue blocks.
        Eventually, the residues are up-sampled to retrieve the recovered secret image.
        }
    \label{fig:network}
\end{figure}

The embedding process is briefly described in Alg. \ref{alg:emb} in the form of pseudo codes.
\begin{algorithm}[t!]
    \caption{Secret Embedding}
    \label{alg:emb}


    $q \gets E(s)$; \Comment{Extract the feature map}
    
    $q' \gets f(q, d)$; \Comment{Get the filtered counterpart}
    
    $c' \gets c + q$; \Comment{Get the container by direct addition}

    Compute $l_{\text{emb}}$ to guarantee the secret embedding\;
    
    Compute $l_{\text{freq}}$ to make $E$ extract the low-frequency feature map.
\end{algorithm}

\paragraph{Structure of the Embedding Network.}
For the embedding network depicted in Fig. \ref{fig:network}, we employ a UNet-structure \cite{ronneberger2015u} framework that is believed to have exceptionally excellent performance with respect to translation-based and semantic segmentation tasks.

As depicted in the left side of Fig. \ref{fig:network}, the network structure is symmetric with down-sampling and up-sampling blocks, and notably, features extracted by the down-sampling layers are then passed to the up-sampling layers for further uses.
The down-sampling block consists of a $4 \times 4$ \textit{Conv2d} layer with a stride of 2, a \textit{LeakyReLU} layer with a negative sloop of 0.2, and a \textit{BatchNorm} layer.
A cover image is down-sampled to a one-dimensional vector that is thereafter up-sampled to get $q$.
The up-sampling block includes a $4 \times 4$ \textit{ConvTransposed2d} with a stride of 2, a \textit{ReLU} layer, and a \textit{BatchNorm} layer.
At the end of the up-sampling, a sigmoid function is going to output the $q$ that is ready for embedding.
This structure allows the up-sampling blocks to share the features derived from the down-sampling blocks, and therefore, the information of the input can be preserved for the generation of the output.

\subsection{Retrieval}

\paragraph{Robustness Enhancement via the Attack Layer.}
To further boost the robustness of LIDS, we insert an attack layer $A$ between $E$ and $R$.
The attack layer can launches various types of attacks, such as JPEG compression, noise addition, Gaussian blurring, and low-pass filtering, on $c'$ to yield a damaged container image $c'_a$, defined as
\begin{equation}
    \begin{aligned}
        c'_a = A(c').
    \end{aligned}
\end{equation}
Each attack has a $0.25$ probability to activate, which means that multiple attacks can be performed simultaneously on the same $c'$.
Nevertheless, it is also possible that no attack is launched on $c'$.
Thereafter, $c'_a$ is passed to $R$ for retrieval.

\paragraph{Retrieval of the Secret Image.}
Finally, we employ the retrieval network $R$ to retrieve the recovered secret image $s'$ from $c'_a$, defined as
\begin{equation}
    \begin{aligned}
        s' = R(c'_a).
    \end{aligned}
\end{equation}
We compute the retrieval loss to ensure that the recovered secret image is close to the original secret image, defined as
\begin{equation}
    \begin{aligned}
        l_{\text{ret}} = \frac{1}{CWH} \sum_i^C \sum_j^W \sum_k^H
                        (s_{ijk} - s'_{ijk})^2.
    \end{aligned}
\end{equation}
Similar to Eq. (\ref{eq:l_emb}), this is also a pixel-wise distance measurement between the secret image and the recovered secret image.

\paragraph{Specificity.}
To avoid that $R$ overfits $s$, such that $R$ outputs $s'$ that is identical to $s$ regardless of what it receives, we need to make $R$ able to differentiate the container image from the clean image.
Thus, we make $R$ output a null image, i.e., a pure black image when it receives the clean image.
This is achieved by the following clean loss term:
\begin{equation}
    \begin{aligned}
        l_{\text{cln}} = \frac{1}{CWH} \sum_i^C \sum_j^W \sum_k^H
                        (R(c)_{ijk} - s^N_{ijk})^2,
    \end{aligned}
\end{equation}
where $R(c)$ denotes the output of $R$ given a clean image $c$; $s^N$ represents the null image.

The retrieval process is concisely demonstrated in Alg. \ref{alg:ret} as pseudo codes.
\begin{algorithm}[t!]
    \caption{Secret Retrieval}
    \label{alg:ret}

    $c_a' \gets A(c')$; \Comment{Attack the container}
    
    $s' \gets R(c_a')$; \Comment{Retrieve the recovered secret}
    
    $\hat{s}^N \gets R(c)$; \Comment{Null Retrieval}
    
    Compute $l_{\text{ret}}$ to guarantee the secret retrieval\;

    Compute $l_{\text{cln}}$ to improve the specificity.
\end{algorithm}

\paragraph{Structure of the Retrieval Network.}
For the retrieval network, we employ a CEILNet-structure \cite{fan2017generic} framework that is believed to function well when the inputs and outputs are distinct.

This network comprises three parts, which are the down-sampling part, the residue part, and the up-sampling part.
In the down-sampling part, the unit block consists of a $3 \times 3$ \textit{Conv2d} layer, a \textit{BatchNorm} layer, and a \textit{ReLU} layer.
Specifically, in the last block of the down-sampling part, the stride is set to two in order to enlarge the receptive field.
In the residue part, there are in total nine residue blocks, which are $18$ convolution layers, to generate residues for further use.
Lastly, in the up-sampling part, the residues are up-sampled to retrieve the recovered secret image $s'$.
The unit block here is similar to that in the down-sampling part, only with differences in the number of input and output channels.
For up-sampling, the first block's \textit{Conv2d} layer's stride and kernel size is set to $2$ and $4 \times 4$.
Eventually, a sigmoid function gives the final $s'$.
CEILNet has a considerably high performance when its output differs from its input significantly, and it is thus suitable for retrieving the recovered secret image from the container image.

\subsection{Loss Function}

The total loss comprises four terms, which are the embedding loss $l_{\text{emb}}$, the frequency loss $l_{\text{freq}}$, the retrieval loss $l_{\text{ret}}$, and the clean loss $l_{\text{cln}}$, defined as
\begin{equation}
    \begin{aligned}
        \mcl L = \lambda_1 l_{\text{emb}} +
                 \lambda_2 l_{\text{freq}} +
                 \lambda_3 l_{\text{ret}} +
                 \lambda_4 l_{\text{cln}},
    \end{aligned}   
\end{equation}
where each $\lambda$ is the weight parameter for each loss term.

$l_{\text{emb}}$ measures the pixel-wise distance $c$ and $c'$, guaranteeing the similarity the two images.
$l_{\text{freq}}$ regulates $q$, such that the frequency distribution of $q$ is close to $q'$ filtered $f$.
By adjusting $d$, one can control the frequency distribution of the secret embedding process.
Similar to $l_{\text{emb}}$, $l_{\text{ret}}$ measures the pixel-wise distance between $s$ and $s'$, ensuring the quality of the retrieval process.
Lastly, $l_{\text{cln}}$ is designed to make $R$ able to differentiate $c'$ from $c$ so as to avoid overfitting.

\section{Experiment}
\label{sec:exp}

\begin{table*}[ht!]
  \centering
  \scriptsize
  \renewcommand\arraystretch{1.5}
  \setlength{\tabcolsep}{1.1mm}{

    \caption{
        Experiment Results: Fidelity, Robustness and Specificity.
        $\uparrow$ indicates that the greater the value is, the better the attribute is, whereas $\downarrow$ means the opposite.
        LIDS has the best performance among the methods with a slight cost of the recovered secret image's quality.
        The cost has earned greatly enhanced robustness in return.
        }
    \label{tab:exp_res}%
    \begin{tabular}{c|cccc|cccccccccc|cccc}
        \Xhline{0.3ex}
        
        \multirow{3}{*}{\textbf{Method}} & \multicolumn{4}{c|}{\textbf{Fidelity}} & %
            \multicolumn{10}{c|}{\textbf{Robustness}} & \multicolumn{4}{c}{\textbf{Specificity}}\\
        \cline{2-19}
         & \multicolumn{2}{c|}{Container} & \multicolumn{2}{c|}{Recovered Secret} & \multicolumn{2}{c|}{JPEG} & \multicolumn{2}{c|}{Low-pass} & \multicolumn{2}{c|}{Blur} & \multicolumn{2}{c|}{Colour Jitter} & \multicolumn{2}{c|}{Resize Crop} & \multicolumn{2}{c|}{Clean} & \multicolumn{2}{c}{Damaged Clean}\\
         & PSNR$\uparrow$ & \multicolumn{1}{c|}{SSIM$\uparrow$} & PSNR$\uparrow$ & SSIM$\uparrow$ & NCC$\uparrow$ & \multicolumn{1}{c|}{SR$\uparrow$} & NCC$\uparrow$ & \multicolumn{1}{c|}{SR$\uparrow$} & NCC$\uparrow$ & \multicolumn{1}{c|}{SR$\uparrow$} & NCC$\uparrow$ & \multicolumn{1}{c|}{SR$\uparrow$} & NCC$\uparrow$ & SR$\uparrow$ & NCC$\downarrow$ & \multicolumn{1}{c|}{SR$\downarrow$} & NCC$\downarrow$ & SR$\downarrow$\\
         
        \Xhline{0.3ex}
        
        \textbf{LIDS} & \textbf{46.60} & \textbf{0.9616} & 32.16 & 0.8277 & \textbf{0.962} & \textbf{91.7\%} & \textbf{0.967} & \textbf{92.6\%} & \textbf{0.986} & \textbf{97.3\%} & \textbf{0.844} & \textbf{71.6\%} & \textbf{0.843} & \textbf{18.6\%} & \textbf{0.428} & \textbf{0.0\%} & \textbf{0.447} & \textbf{0.0\%}\\
        
        UDH & 26.24 & 0.7778 & 22.11 & 0.6884 & 0.580 & 0.0\% & 0.629 & 0.0\% & 0.629 & 0.0\% & 0.491 & 2.2\% & 0.685 & 0.3\% & 0.629 & 0.0\% & 0.629 & 0.0\%\\
        
        DDH & 31.08 & 0.7128 & 34.65 & 0.7813 & 0.675 & 0.0\% & 0.878 & 0.0\% & 0.912 & 0.0\% & 0.452 & 0.8\% & 0.763 & 1.1\% & 0.441 & 0.0\% & 0.491 & 0.0\%\\
        
        HiNet & 32.77 & 0.6984 & \textbf{35.46} & \textbf{0.8529} & 0.620 & 0.0\% & 0.616 & 0.0\% & 0.619 & 0.0\% & 0.884 & 65.4\% & 0.592 & 0.3\% & 0.594 & 0.0\% & 0.611 & 0.0\%\\




        
        \Xhline{0.3ex}
    \end{tabular}
    }
\end{table*}%

\subsection{Implementation Details}

\paragraph{Experiment Settings.}
We select FFHQ \cite{karras2019style} and CelebA-HQ \cite{liu2018large} datasets to be the training and test cover image sets.
These datasets have various and subtle features, and we thus believe that they are capable to prove the generalizability of our method.
The images are cropped and resized into size of $256 \times 256$ to fit the input size of the network.
For evaluation, we randomly select $1,000$ images from the CelebA-HQ dataset to form the evaluation set.

We use the Adam optimizer to train LIDS with a learning rate of $0.0001$ and betas of $(0.1, 0.5)$ as the initial value.
In our warm-up experiments, it is proved that increasing the learning rate and betas does not hasten convergence, but rather impede it.
The batch size is set to $16$, and the number of training steps in one epoch is limited to 2500.
The model is trained for $12$ epochs, and we decrease the learning rate by $0.2$ every three epochs.
$d$ is set to $50$ by default.

\paragraph{Evaluation Metrics.}

Peak signal-to-noise ratio (PSNR) and structural similarity index measure (SSIM) \cite{hore2010image} were used to measure the fidelity.
Here, high fidelity indicates a low visual distance of the two image pairs, which are the cover image and the container image, and the secret image and the recovered secret image.
In this context, higher averaged PSNR and SSIM values refer to better fidelity.

Given an original image $x$ and its noisy approximation $x'$ of the same size $m \times n$, PSNR is defined as
\begin{equation}
    \begin{aligned}
        \text{PSNR}(x, x') = & \ 20 \cdot \log_{10} (\max(x)) -\\
                      & \ 10 \cdot \log_{10} (\text{MSE}(x, x')),
    \end{aligned}
\end{equation}
where $\max(x)$ denotes the maximum possible pixel value of $x$.
For example, if the pixels are encoded in 8-bit format, $\max(x) = 255$.
The $MSE$ operator is further defined as 
\begin{equation}
    \begin{aligned}
        \text{MSE}(x, x') = \frac{1}{mn} \sum_i^m \sum_j^n \| x_{i, j} - x'_{i, j} \|^2.
    \end{aligned}
\end{equation}

Given two images $x, y$ of the same size $n \times n$, SSIM is defined as
\begin{equation}
    \begin{aligned}
        \text{SSIM}(x, y) = \frac{(2 \mu_x \mu_y + c_1) (2 \sigma_{xy} + c_2)}
                          {(\mu_x^2 + \mu_y^2 + c_1) (\sigma_x^2 + \sigma_y^2 + c_2)},
    \end{aligned}
\end{equation}
where $c_1 = (k_1 L)^2$ and $c_2 = (k_2 L)^2$.
$L$ is the dynamic range of the pixel values, e.g., if an image is encoded in 8-bit format, $L = 2^8 - 1 = 255$.
$k_1, k_2$ are two constants, and set to $0.01$ and $0.03$ by default.

The robustness of the methods are evaluated by Normalized Cross Correlation (NCC) and Success Rate (SR) \cite{zhang2021deep}.
Here, robustness is defined as the ability to retrieve a valid recovered secret image from the damaged container image.
To determine whether a recovered secret image is valid or not, we employed NCC.
Given a secret image $s$ and a recovered secret image $s'$, NCC is defined as
\begin{equation}
    \begin{aligned}
        \text{NCC} = \frac{\langle s', s \rangle}{\| s' \| \cdot \| s \|}.
    \end{aligned}
\end{equation}
Here, $\langle \cdot , \cdot \rangle$ denotes the inner product, and $\| \cdot \|$ denotes the L2 norm.
Briefly, NCC measures the similarity between two images.
The larger the NCC is, the more similar the two images are.
We consider an NCC greater than $0.95$ as a symbol of a valid secret retrieval.

In order to further evaluate an IDS method's robustness, we define SR as
\begin{equation}
    \begin{aligned}
        \text{SR} = \frac{\text{\# Valid Secret Retrieval}}{\text{\# Secret Retrieval}}.
    \end{aligned}
\end{equation}
This is the ratio of the number of valid secret retrieval to the total number of secret retrieval.
Higher averaged NCC and SR indicate superior robustness.

Specificity is also evaluated by NCC and SR, but it is different from the robustness evaluation.
Specificity is defined as the retrieval network's ability to differentiate the container image from the clean image.
In other word, when the retrieval network receives a clean image that does not have any hidden content, the retrieval network should output a null image, i.e., a pure black image.
Thus, when we compute the NCC and SR using the secret image and the recovered null image, lower NCC and SR indicate better specificity in contrast.

\begin{figure*}[ht!]
    \centering
    \includegraphics[width=\textwidth]{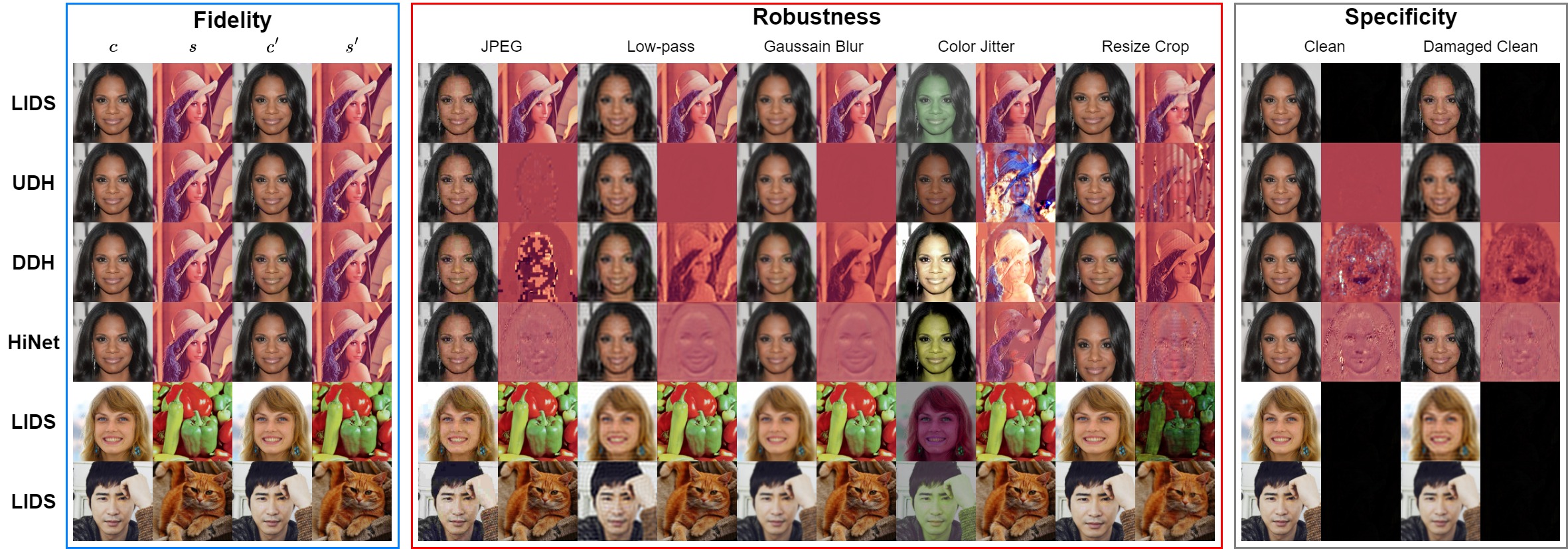}
    \caption{
        Experiment Results - Fidelity, Robustness \& Specificity.
        The top four rows are the comparison experiment results that show LIDS has the best robustness against different types of attacks while maintaining high fidelity and specificity.
        Meanwhile, LIDS's performance remains when facing different secret images shown in the last two rows.
        }
    \label{fig:exp_res}
\end{figure*}

\subsection{Comparison of Fidelity, Robustness, and Specificity}
To validate the effectiveness of our method, we trained several IDS models with the same datasets.
The models are Universal Deep Hiding (UDH) \cite{zhang2020udh} by Zhang et al., HiNet by \cite{jing2021hinet}, and Dependent Deep Hiding (DDH) \cite{weng2019high}, which are considered exemplary works of IDS.
These models are trained using their default settings, and validated with the evaluation set.

\paragraph{Comparison of fidelity.}
To compare the fidelity of different IDS methods, we first generate $1,000$ container images using each of them with the evaluation set.
Then, we compute the averaged PSNR and SSIM of each group of container images and recovered secret images retrieved from the container images.
The results are illustrated in Tab. \ref{tab:exp_res}'s ``Fidelity" Column.
Fig. \ref{fig:exp_res}'s ``Fidelity" section (blue) shows the container images and recovered secret images generated by each of the methods.

LIDS has outperformed all other methods in terms of the fidelity of the container image, as evidenced by its superior PSNR and SSIM values.
Nevertheless, HiNet and DDH have yielded superior results in terms of the recovered secret image.
The inability of LIDS to retrieve the recovered secret image as well as the other two models may be due to the fact that the extracted features were primarily concentrated in the low-frequency domain.
However, this does not imply that LIDS was outperformed by the others, as the visual quality of the container images weighs significantly more than the fidelity of the recovered secret images.
The parameters are in the supplementary materials.

\paragraph{Comparison of robustness.}
To further test each of the methods' robustness against malicious attacks, we launch various types of attacks on each group of the container images, including JPEG compression, low-pass filtering, Gaussian blur, colour jitter, and resize and cropping, to yield damaged container images.Each group of the damaged container images is then fed into the corresponding retrieval network for retrieval.
We then compute the success rate of the secret retrieval for each of the networks so as to compare their robustness against attacks, and the results are shown in Tab. \ref{tab:exp_res}'s ``Robustness" column.
The performance of each IDS method against different type of attacks is depicted in Fig. \ref{fig:exp_res}'s ``Robustness" section (red).

In terms of robustness, LIDS has exceeded the other three methods.
LIDS is extremely resistant to attacks that severely distorted the high-frequency signal, such as JPEG compression, low-pass filtering, and Gaussian blurring.
The color jitter attack drastically alters the color distribution of the images, thereby posing a threat to the container images' low-frequency component.
The resizing and cropping attack is deemed the most effective method of attack against the embedded secret with a non-negligible visual quality cost, because it eliminates the embedded secret signal by geometrically reshaping the entire image.
Nonetheless, it is quite encouraging to see that LIDS is able to partially withstand these two powerful attacks while the others fail.

Notably, the other three models are able to attain high NCC values despite having pretty low SRs.
Upon examining the recovered images, we discover that the networks are able to retrieve invalid secret images that are filled with a red hue.
Due to the absence of the clean loss term $l_{\text{cln}}$, these networks may overfit to the secret image, thereby retrieving the invalid secret image.

\paragraph{Comparison of specificity.}

In Fig. \ref{fig:exp_res}'s ``Specificity" section (gray), LIDS produces the null images when it receives the clean images, whereas the rest of the models yield images with noise to an extent.
The numeric results are listed in Tab. \ref{tab:exp_res}'s ``Specificity" column, where LIDS has the best performance.
This is accomplished by training the model using the clean loss term $l_{\text{cln}}$.

To further exclude the possibility that LIDS could retrieve valid secret images from damaged clean images by recognizing attack patterns, we generate damaged clean images from the clean images.
The damaged images are then used to evaluate the models' specificity.
As a result, LIDS is able to easily distinguish between the damaged clean images and the container images, producing black outputs while the other models performed similarly as 
they did when presented with the clean images.

\subsection{Comparison with the SOTAs}
\begin{table}[t!]
  \centering
  \scriptsize
  \renewcommand\arraystretch{1.2}
  \setlength{\tabcolsep}{2.7mm}{
    \caption{
        Robustness Comparison with the SOTAs (PSNR).
        The higher the PSNR value under an attack is, the better robustness the method has.
        LIDS has the best robustness among these methods in terms of the recovered secret images.
        }
    \label{tab:sotas}%
    \begin{tabular}{c|cc|c|cc}
        \Xhline{0.3ex}
        
        \multirow{2}{*}{\textbf{Method}} & \multicolumn{2}{c|}{Gaussian Noise} & 
            \multicolumn{1}{c|}{Poisson}  & \multicolumn{2}{c}{JPEG}\\
         & $\sigma = 10$ & \multicolumn{1}{c|}{$\sigma = 1$} & \multicolumn{1}{c|}{Noise} & QF $=40$ & QF $=90$\\
         
        \hline
        
        ISN & 8.55 & 25.19 & 19.38 & 10.11 & 11.25\\
        ISN$^+$ & 27.12 & 28.98 & 26.71 & 26.25 & 27.48\\
        RIIS$^*$ & 28.03 & 30.01 & 27.23 & 27.18 & 28.44\\
        RIIS & 28.22 & 30.32 & 27.47 & 27.32 & 28.71\\
        \textbf{LIDS} & \textbf{29.03} & \textbf{30.51} & \textbf{28.31} & \textbf{30.71} & \textbf{32.14}\\
        
        \Xhline{0.3ex}
    \end{tabular}
    }
\end{table}%

ISN \cite{lu2021large} and RIIS \cite{xu2022robust} are currently the state-of-the-art IDS methods, and thus, we make a robustness comparison with these two methods.
Due to the fact that there is no official released implementation of the methods, we directly use the experiment results in \cite{xu2022robust}, and produce our results under the same conditions.
We use the DIV2k \cite{Agustsson_2017_CVPR_Workshops} dataset to train and evaluate LIDS.

As shown in Tab. \ref{tab:sotas}, we launch three types of attacks with different parameters on the container images, and retrieve the recovered secret images with LIDS.
ISN$^+$ is the fine-tuned version of ISN appeared in the experiments in \cite{xu2022robust}, and has achieved enhanced robustness compared to ISN.
RIIS is another version of RIIS that has a unified framework for all distortion built up with distortion-guided modulation.
LIDS has the highest averaged PSNR values of the recovered secret images compared to the others, which indicates that LIDS has the best robustness compared to the state-of-the-art IDS methods.

\section{Analysis}
\label{sec:analysis}

\begin{table*}[t]
  \centering
  \scriptsize
  \renewcommand\arraystretch{1.5}
  \setlength{\tabcolsep}{1.1mm}{
    \caption{
        Ablation Study.
        $\uparrow$ indicates that the greater the value is, the better the attribute is, whereas $\downarrow$ means the opposite.
        Without the attack layer or the frequency loss, LIDS has a significant boost in fidelity but with an obvious drop in robustness.
        Without the clean loss term, LIDS overfits the secret image, and thus cannot differentiate the clean images from the container images.
        }
    \label{tab:ablation_total}%
    \begin{tabular}{ccc|cc|ccccc|cc}
        \Xhline{0.3ex}
        
        \multicolumn{3}{c|}{\textbf{Building Blocks}} & \multicolumn{2}{c|}{\textbf{Fidelity}} & %
            \multicolumn{5}{c|}{\textbf{Robustness}} & \multicolumn{2}{c}{\textbf{Specificity}}\\
            
        \cline{1-12}
        
         \multicolumn{1}{c|}{Attack} & \multicolumn{1}{c|}{Frequency} & \multicolumn{1}{c|}{Clean} & \multicolumn{1}{c|}{Container} & \multicolumn{1}{c|}{Recovered Secret} & \multicolumn{1}{c|}{JPEG} & \multicolumn{1}{c|}{Low-pass} & \multicolumn{1}{c|}{Blur} & \multicolumn{1}{c|}{Colour Jitter} & \multicolumn{1}{c|}{Resize Crop} & \multicolumn{1}{c|}{Clean} & Damaged Clean\\
         
        
         \multicolumn{1}{c|}{Layer} & \multicolumn{1}{c|}{Loss} & \multicolumn{1}{c|}{Loss} & \multicolumn{1}{c|}{PSNR/SSIM$\uparrow$} & \multicolumn{1}{c|}{PSNR/SSIM$\uparrow$} & \multicolumn{1}{c|}{NCC/SR$\uparrow$} & \multicolumn{1}{c|}{NCC/SR$\uparrow$} & \multicolumn{1}{c|}{NCC/SR$\uparrow$} & \multicolumn{1}{c|}{NCC/SR$\uparrow$} & \multicolumn{1}{c|}{NCC/SR$\uparrow$} & \multicolumn{1}{c|}{NCC/SR$\downarrow$} & NCC/SR$\downarrow$\\

        \Xhline{0.3ex}
        
        $\times$ & $\checkmark$ & $\checkmark$ & 51.47/0.9632 & 46.32/0.9666 & 0.333/20.0\% & 0.998/99.7\% & 0.945/87.3\% & 0.701/60.1\% & 0.571/3.7\% & 0.271/0.0\% & 0.255/0.0\%\\
        
        $\checkmark$ & $\times$ & $\checkmark$ & 46.64/0.9950 & 44.64/0.9365 & 0.904/87.8\% & 0.790/69.8\% & 0.932/93.6\% & 0.844/68.5\% & 0.533/10.6\% & 0.002/0.0\% & 0.040/0.0\%\\

        $\checkmark$ & $\checkmark$ & $\times$ & 42.73/0.9814 & 52.37/0.9999 & 0.995/100\% & 0.999/100\% & 0.999/100\% & 0.997/100\% & 0.989/100\% & 0.981/100\% & 0.989/100\%\\

        $\checkmark$ & $\checkmark$ & $\checkmark$ & 42.82/0.9806 & 34.23/0.8506 & 0.987/95.1\% & 0.949/81.4\% & 0.995/99.1\% & 0.935/77.4\% & 0.602/17.4\% & 0.281/0.0\% & 0.294/0.0\%\\
        
        \Xhline{0.3ex}
    \end{tabular}
    }
\end{table*}%

\subsection{Residue Analysis}

\begin{figure}[t!]
    \centering
    \includegraphics[width=.49\textwidth]{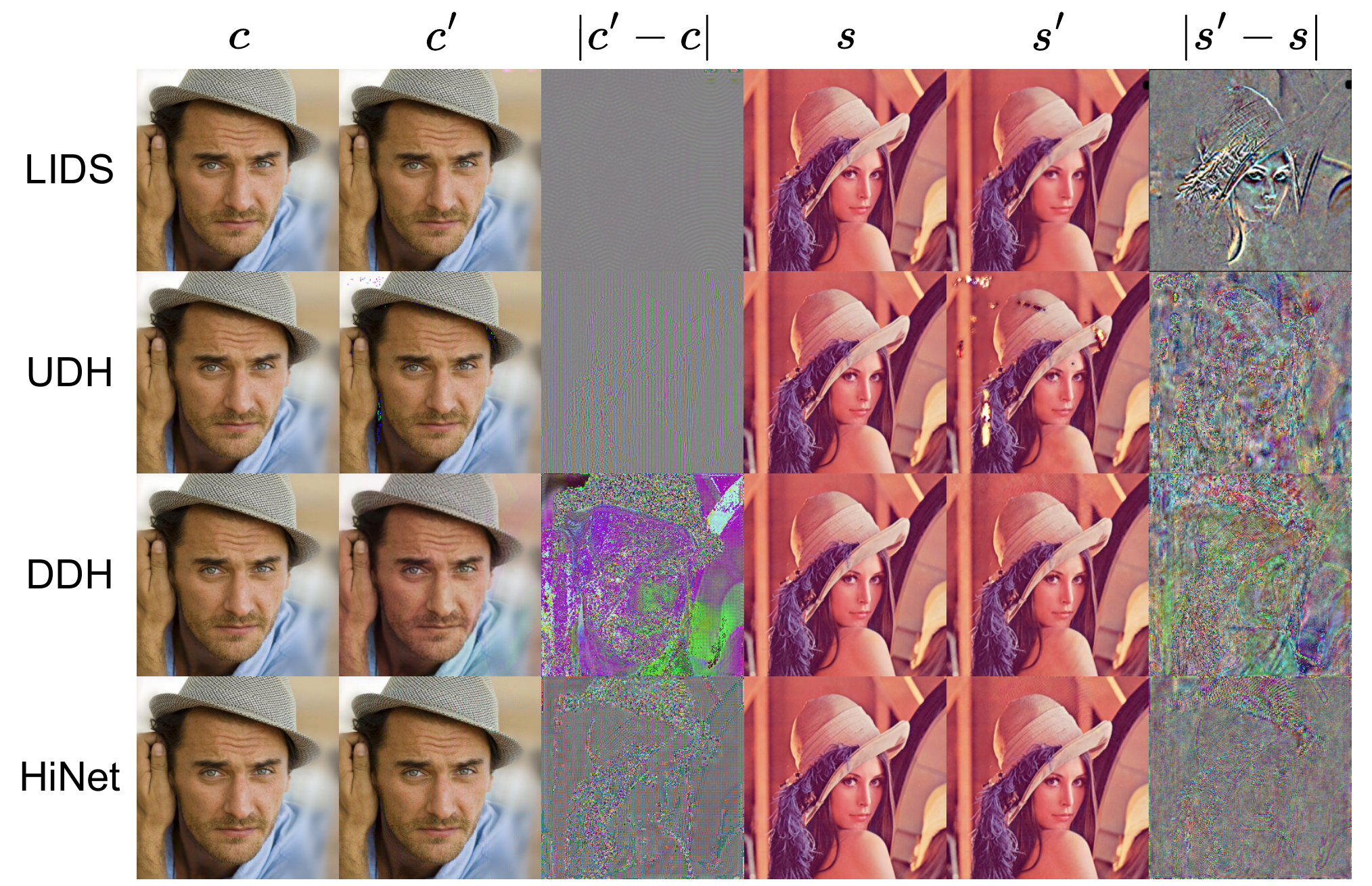}
    \caption{
        Residues Analysis (10$\times$ Enhanced).
        Visually, LIDS has the least residues between the cover and the container image.
        The residues between the secret and the recovered secret image of LIDS show a profile of the image edge, which is the image's high-frequency components.
        }
    \label{fig:residues}
\end{figure}

To further investigate how the IDS methods work, we perform the residue analysis by subtracting $c$ from $c'$, and $s$ from $s'$.
The residues are enhanced 10 times so as to reveal the tiny differences therein as depicted in Fig. \ref{fig:residues}.

Regardless of the method, we can conclude that the model embeds $s$ by altering the pixel values of $c'$.
However, each method embeds $s$ using a distinct strategy.
LIDS has the fewest residues between $c$ and $c'$, resulting in the highest PSNR and SSIM values among the container images in Tab. \ref{tab:exp_res}.
LIDS's residues between $s$ and $s'$ exhibit a distinct profile of $s$, which may account for the inferior visual quality of $s'$ in comparison to the others'.
The high-frequency component of an image contains the majority of the image's profile information; therefore, if only the low-frequency features are extracted, the recovered image will lack the high-frequency component.
Eventually, the visual quality of LIDS's recovered secret image is inferior to that of the others due to the low-frequency embedding process, but only if no attack is launched on $c'$.

\subsection{Frequency Analysis}
\begin{figure}[t!]
    \centering
    \includegraphics[width=.49\textwidth]{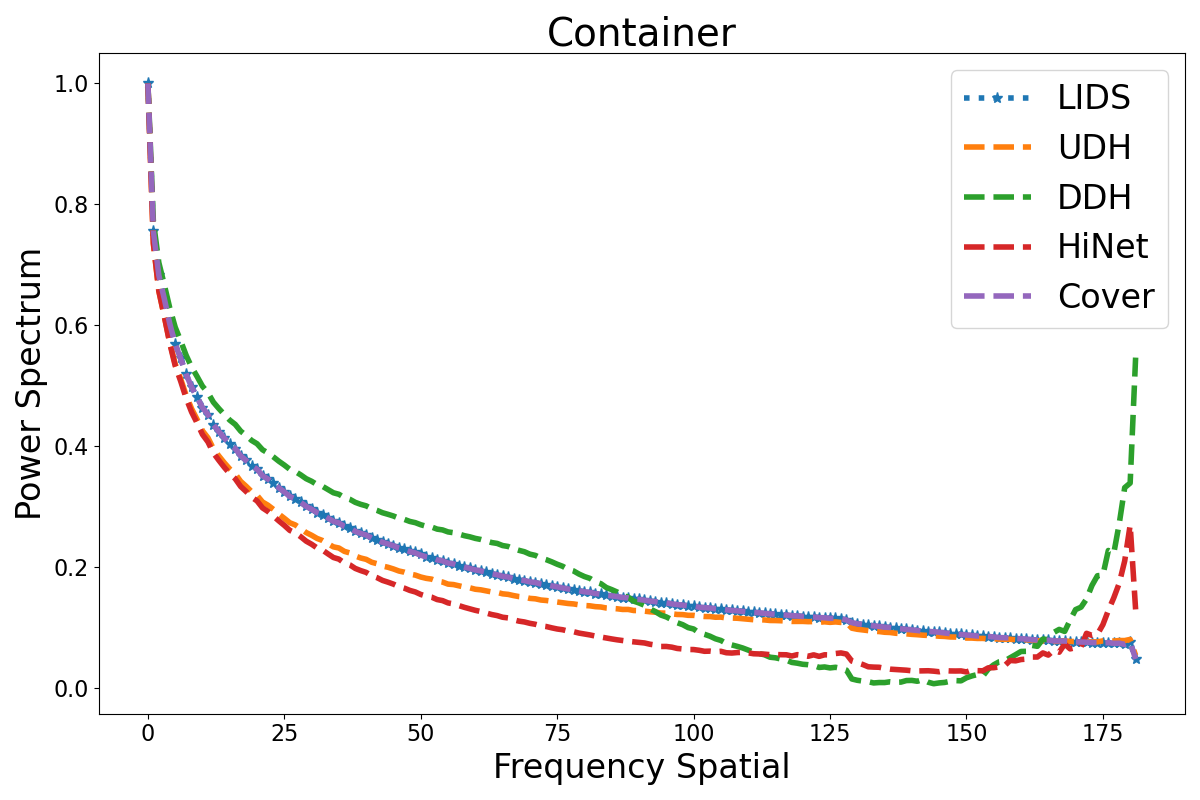}
    \caption{
    Frequency analysis of the container images.
        The perfect result is that the frequency distribution of the container images is identical to that of the cover images (purple dashed line).
        The averaged frequency spectrum of $1,000$ container images generated by LIDS (blue dashed line marked with stars) aligns almost perfectly to that of the cover images.
        }
    \label{fig:freq_als_ctn}
\end{figure}

\begin{figure}[t!]
    \centering
    \includegraphics[width=.49\textwidth]{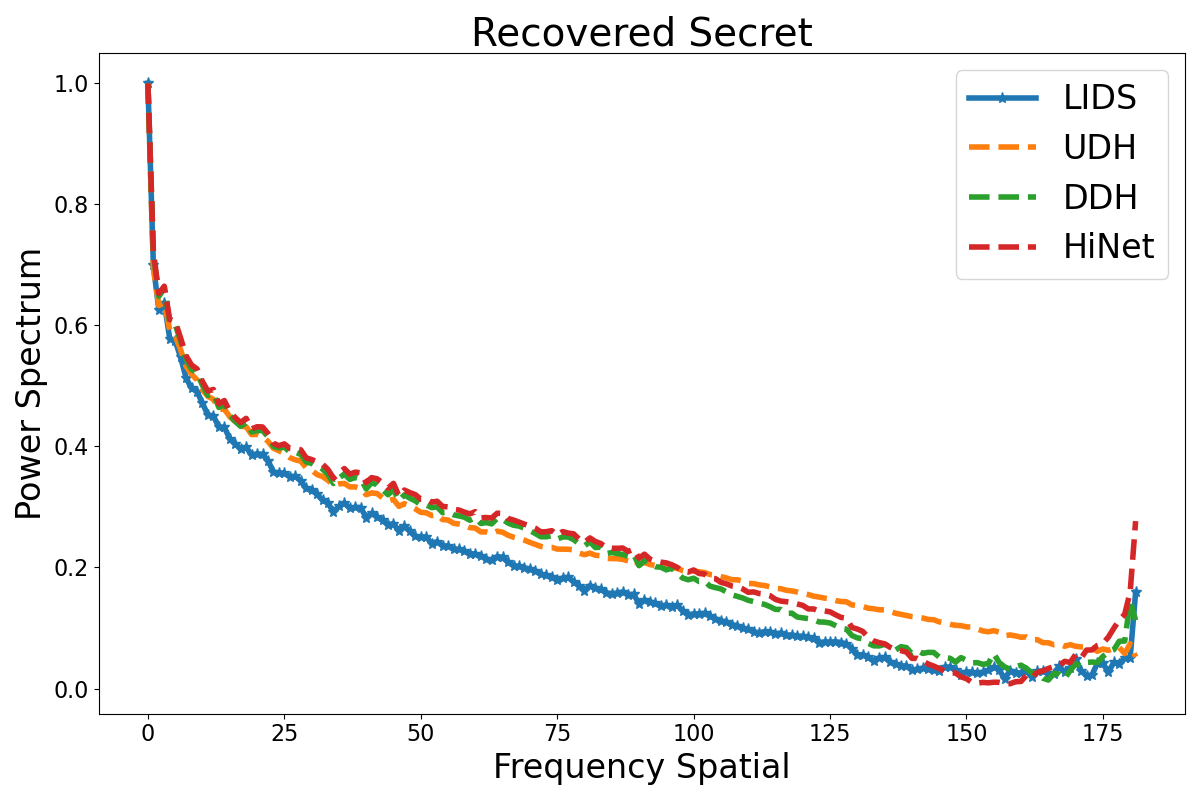}
    \caption{
        Frequency analysis of the secret images.
        As expected, the frequency distribution of the recovered secret images of LIDS should have less high-frequency information.
        Compared to the averaged frequency spectrum of $1,000$ recovered secret images by the other methods, that of LIDS meets the expectation.
        }
    \label{fig:freq_als_sc}
\end{figure}

To make a comparison in the frequency domain, we computed the averaged Azimuthal Integral \cite{durall2020watch} of container and secret images generated by the models.

In brief, Azimuthal Integral computes the radial integral over the 2D discrete Fourier Transform spectrum along the spatial frequency.
Given a square image $I$ of size $M \times N$($M = N$), the spectral representation is computed from the discrete Fourier Transform (DCT)
\begin{equation}
    \begin{aligned}
        & \text{DCT}(I)(k, l) = \sum_{m=1}^{M} \sum_{n=1}^{N}
                              e^{-2\pi i \cdot \frac{jk}{M}}
                              e^{-2\pi i \cdot \frac{jl}{N}}
                              \cdot I(m, n),\\
        & \text{for} \quad k=1,...,M, \quad l=1,...,N,
    \end{aligned}
\end{equation}
via Azimuthal Integration over radial frequencies $\phi$
\begin{equation}
    \begin{aligned}
        & \text{AI}(\omega_k) = \int_0^{2\pi} \| 
                              \text{DCT}(I) 
                              (\omega_k \cdot \cos(\phi),
                              \omega_k \cdot \sin(\phi)) 
                              \|^2 d\phi\\
        & \text{for} \quad k=1, ..., M/2.
    \end{aligned}
\end{equation}

\begin{figure}[t]
    \centering
    \includegraphics[width=.47\textwidth]{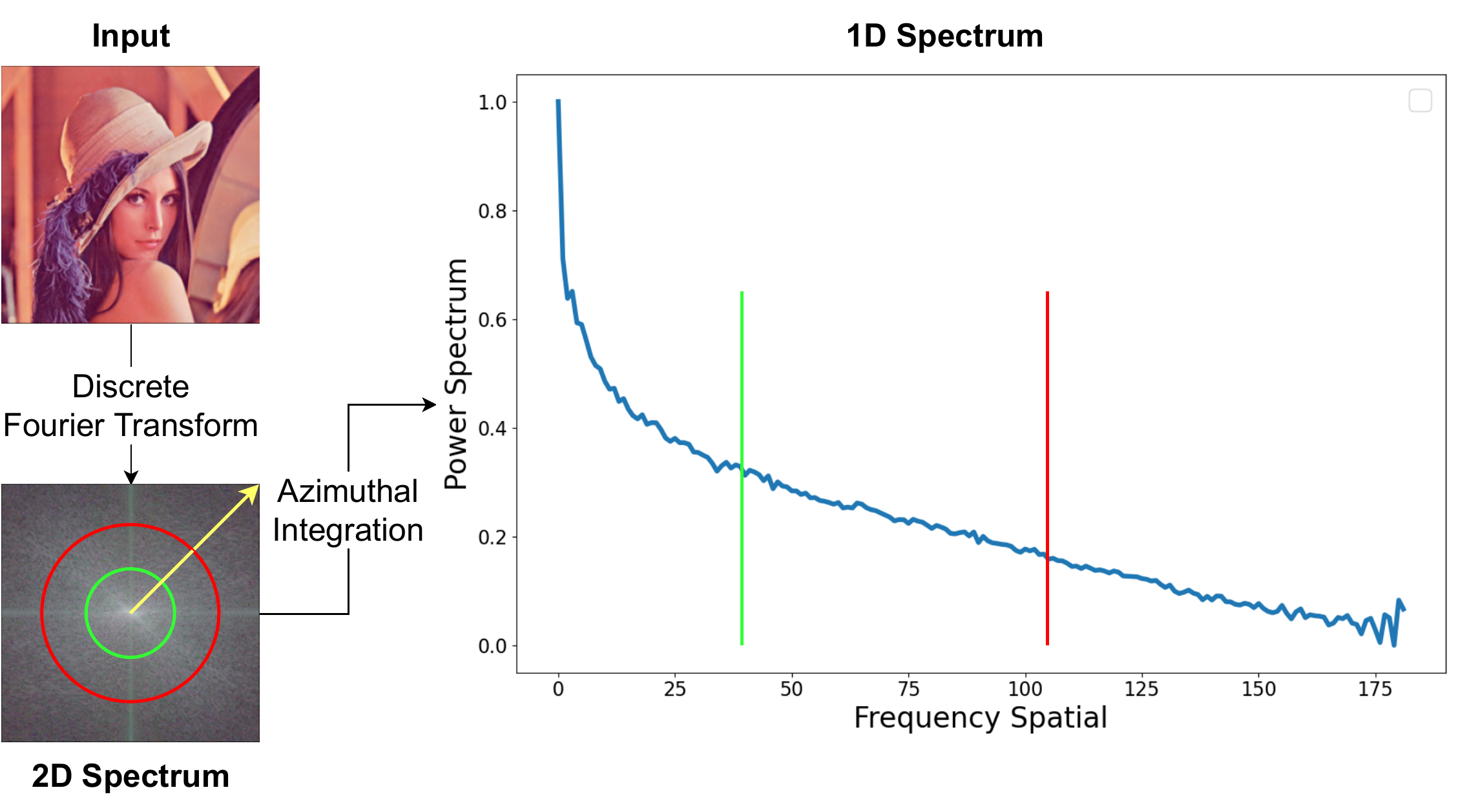}
    \caption{
            Azimuthal Integral.
            The input image is transformed into a 2D spectrum using discrete Fourier transform.
            Then, the integration is conducted from the inside of the 2D spectrum to the outside, along the yellow arrow circularly.
            Eventually, the 1D spectrum is derived, where the red and green lines represent the sum of the pixel values on the red and green circles.
            }
    \label{fig:azi_int}
\end{figure}

As depicted in Fig. \ref{fig:azi_int}, the 1D Azimuthal Integral power spectrum reflects the relative intensity of the 2D spectrum at a certain frequency spatial coordinate.

In Fig. \ref{fig:freq_als_ctn}, we can clearly see that the frequency distribution of LIDS's container images almost perfectly aligns to that of the cover images, which explains why LIDS works well.
As for the others, the tails of the frequency distributions of the container images generated by DDH and HiNet have a significant sharp frequency ascent, making their container images less likely to withstand attacks aiming at frequency domain.
UDH's and LIDS's container images have a similar frequency distribution, but we can easily distinguish between them.

The recovered secret images analysed in Fig. \ref{fig:freq_als_sc} are all directly output by the retrieval networks, as evidenced by their sharp ascent at their tails shown.
As expected, LIDS's recovered secret images contain less high frequency information than that of the others, which explains why LIDS's recovered secret images has lower quality than those generated by the other methods.

Our motivation of conducting IDS in low-frequency domain is not only supported by the experimental results, but also with theoretical analysis .
In \cite{fan2020low}, it is pointed out that the watermark embedded into the image's low-frequency features is more robust, because the low frequency feature is steady.
\cite{daren2001dwt} claims that watermark should be embedded into the low-frequency domain based on a quantitative analysis on magnitude of wavelet coefficients.
\cite{singh2016hybrid} demonstrates that the human eyes are much more sensitive to the low-frequency components of images than the high-frequency components, and the watermarks embedded in the low-frequency components can be easily extracted as long as the information is not lost too much.
\cite{lin2010improving} argues that embedding the watermark into the low-frequency components is a considerable approach to enhance the watermark's robustness, though this method brings undesired degradation to the image's quality.

To further test how LIDS reacts to high-pass filter attack, we conduct an experiment where we extract the embedded secret images from the container images processed by high filter.
In details, we randomly select $1,000$ container images, and respectively launch high/low-pass filter attacks on the container images with various cut-off parameter $d$.
In the high-pass filter attack, greater $d$ indicates that less low-frequency components are kept, whereas in the low-pass filter attack, greater $d$ indicates that more high-frequency components are kept.
Then, we extract the embedded secrets from the container images, and compute the average NCC and SR values.
 
\begin{figure}[t!]
    \centering
    \includegraphics[width=.49\textwidth]{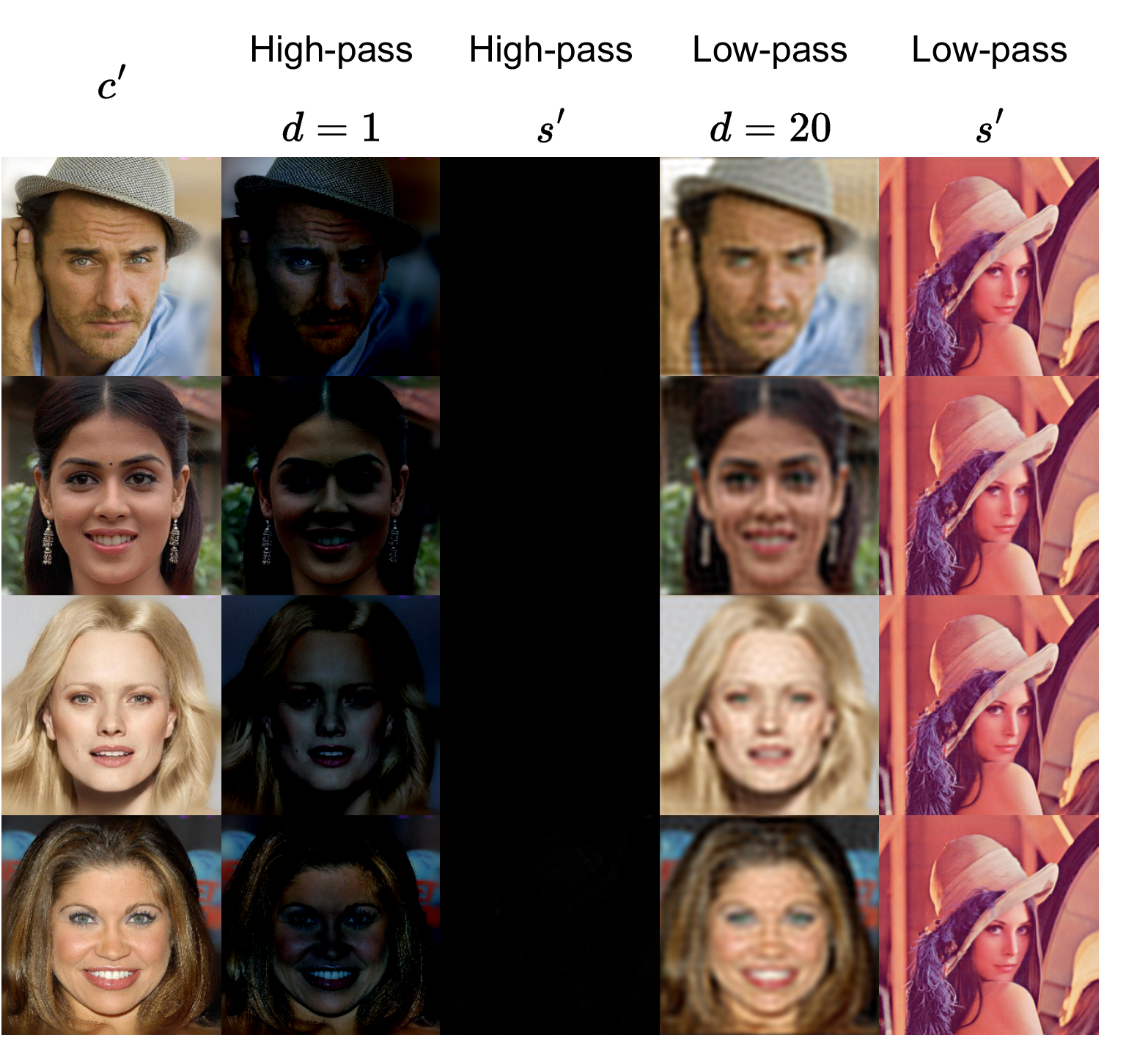}
    \caption{High/low-pass Filter Attack Results:
            Even though high-pass filter can remove the embedded secrets with the least level of attack, the perceptual cost is extremely high, whereas the low-pass filter cannot remove the secrets with a large scale of high-frequency components being removed from the container image.
             }
    \label{fig:freq_atk_res}
\end{figure}

\begin{table}[t!]
  \centering
  \scriptsize
  \renewcommand\arraystretch{1.5}
  \setlength{\tabcolsep}{.9mm}{
    \caption{
        High/low-pass Filter Attack Results (NCC/SR).
        Higher NCC/SR indicates better robustness.
        Although high-pass filter attack can easily remove the embedded secret images, the toll is very high in terms of the perceptual quality of the processed container image.
        }
    \label{tab:freq_atk_res}%
    
    \begin{tabular}{c|ccccc}
        \Xhline{0.3ex}
        
        \textbf{d} & 1 & 20 & 40 & 60 & 80\\
         
        \Xhline{0.3ex}

        High-pass & 0.468/0.0\% & 0.457/0.0\% & 0.485/0.0\% & 0.491/0.0\% & 0.443/0.3\%\\

        Low-pass & 0.455/0.0\% & 0.973/93.7\% & 0.981/97.2\% & 0.991/99.9\% & 0.991/100.0\%\\
        
        \Xhline{0.3ex}
    \end{tabular}
    
    }
    
\end{table}%

Given the results in Fig. \ref{fig:freq_atk_res} and Tab. \ref{tab:freq_atk_res}, even if we perform the lightest high-pass filter attack on the container images with $d = 1$, the attacked images are severely damaged, leading to the failures in secret retrieval. 
However, even if we cut off most of the high-frequency components in the container images with $d = 20$, the secret retrieval's SR is high, and the attacked images remain better perceptual quality compared to those high-pass filtered images.
This indicates that the adversary must pay a high toll in order to remove the hidden content inside the container image.
Based on the experiments and relevant works, we may conclude that low-frequency image deep steganography demonstrates higher robustness than the others because of the method's endogenous property, and the perceptual quality cost is non-negligible for the adversary to remove the hidden secrets.




\subsection{Ablation Study}

\begin{table}[t!]
  \centering
  \scriptsize
  \renewcommand\arraystretch{1.5}
  \setlength{\tabcolsep}{.9mm}{
    \caption{
        Robustness Test of LIDS without the Attack Layer (NCC/SR).
        Higher NCC/SR indicates better robustness.
        LIDS without the attack layer is still more robust than the other methods in terms of secret retrieval.
        }
    \label{tab:ablation_atk_layer}%
    
    \begin{tabular}{c|ccccc}
        \Xhline{0.3ex}
        
        \textbf{Method} & JPEG & Low & Blur & Color & Resize Crop\\
         
        \Xhline{0.3ex}
        
        \textbf{LIDS} & \textbf{0.333/20.0\%} & \textbf{0.998/99.7\%} & \textbf{0.945/87.3\%} & 0.701/60.1\% & \textbf{0.571/3.7\%}\\

        UDH & 0.578/0.0\% & 0.628/0.0\% & 0.629/0.0\% & 0.491/2.2\% & 0.685/0.3\%\\

        DDH & 0.885/0.0\% & 0.909/0.0\% & 0.902/0.0\% & 0.452/0.8\% & 0.763/1.1\%\\

        HiNet & 0.690/0.0\% & 0.617/0.0\% & 0.619/0.0\% & \textbf{0.884/65.4\%} & 0.592/0.3\%\\
        
        \Xhline{0.3ex}
    \end{tabular}
    
    }
    
\end{table}%

In order to examine whether LIDS's framework is correctly designed, we conducted several experiments by knocking out LIDS's kernel building blocks each at a time.
To demonstrate the sufficiency and necessity of LIDS's framework, we train several LIDS models with the same settings and datasets, where LIDS's kernel building blocks are removed one at a time.
We then test the models with alleviated attacks.
The results are demonstrated in Tab. \ref{tab:ablation_total}.

\paragraph{The Attack layer.}

\begin{figure}[t!]
    \centering
    \includegraphics[width=.49\textwidth]{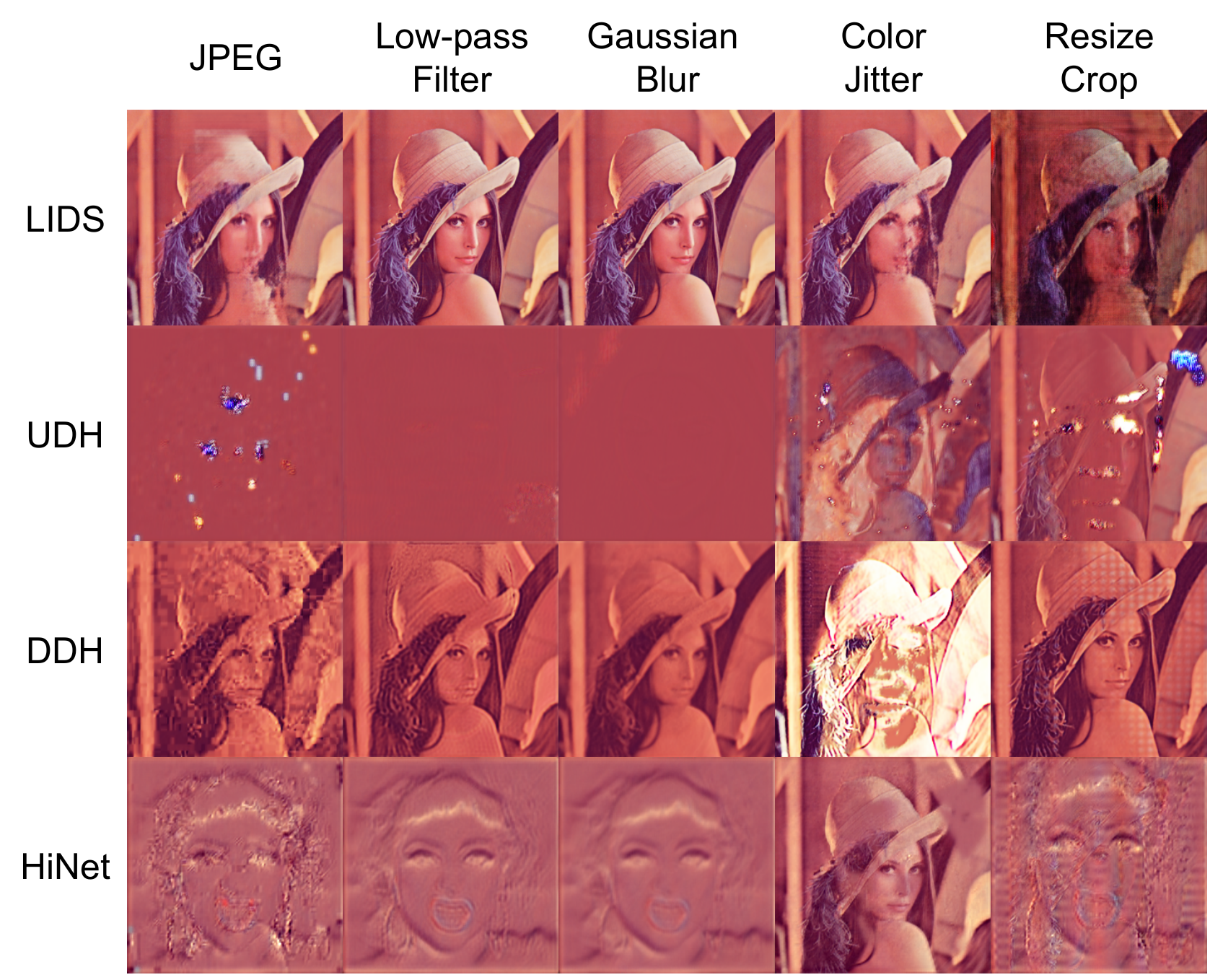}
    \caption{
        Robustness Test of LIDS without the Attack Layer.
        Though the robustness of LIDS decreases significantly without the attack layer, LIDS can still outperform the other methods in terms of secret retrieval.
        }
    \label{fig:ablation_atk_layer}
\end{figure}

Firstly, we remove the attack layer to see the effects.
Judging from the results in Tab. \ref{tab:ablation_total}, the robustness of LIDS drops significantly, whereas the fidelity rises.
This indicates that the attack layer plays an important role in terms of enhancing the robustness, and there is a trade-off between fidelity and robustness when deciding whether to employ the attack layer.

In order to examine whether the low-frequency embedding process is effective, we further conduct a comparison experiment between the LIDS without the attack layer and the other three methods.
The experiment results are listed in Tab. \ref{tab:ablation_atk_layer}.

Even without the attack layer, LIDS outranks the competition in terms of robustness as shown in Fig. \ref{fig:ablation_atk_layer}.
Except for the color jitter attack, LIDS achieved the highest SRs in all other attacks, while its SR against the color jitter attack was considerable.
It demonstrates that LIDS is less effective without the attack layer, but still efficacious.
Remarkably, under the same settings as in those experiment presented in Tab. \ref{tab:exp_res}, the container images and recovered secret images generated by this LIDS model surprisingly reached a extremely high visual quality.
with the PSNR/SSIM values of $(51.47/0.9632, 46.32/ 0.9666)$, respectively.
This suggests that there is a trade-off between high fidelity and high robustness by employing the attack layer or not.
Additionally, it also proves that the attack layer can greatly boost the model's robustness, for the standard LIDS could best all other methods under the extreme attacks.
In addition, this demonstrates that the attack layer can greatly enhance the robustness of the model, as the standard LIDS outscored all other methods when subjected to extreme attacks.

\paragraph{Frequency loss.}
We trained an LIDS model by omitting the frequency loss term to test how it affected the watermarking network's performance.
Surprisingly, we found that the model's $l_{\text{emb}}$ decreased to 0, whereas the model's $l_{\text{ret}}$ and $l_{\text{cln}}$ remained still.
During the training, the embedding network somehow chose not to extract any features from the watermark so as to reduce $l_{\text{emb}}$ to 0, from which the embedding network benefited at most.
Having completely no connection to the embedding network, the retrieval process could thus gain nothing from the drop of $l_{\text{emb}}$.
Moreover, the retrieval network could not differentiate the watermarked images from the clean images, because no feature was added to the watermarked images.
This led to the not converging $l_{\text{ret}}$ and $l_{\text{cln}}$.

The training processes showed random behaviors that the model sometimes did not converge.
We discovered that the model's embedding loss $\mcl L_{\mcl E}$ decreased to 0 while its $l_\text{ret}$ and $l_\text{cln}$ remained unchanged.
During training, the embedding network chose not to extract any features from the secret image in order to reduce $\mcl L_{\mcl E}$ to 0, which was most advantageous to the embedding network.
Having no connection whatsoever to the embedding network, the retrieval process gained no benefit from the drop of $\mcl L_{\mcl E}$.
In addition, the retrieval network was unable to distinguish between the container and clean images because the container images lacked, and consequently, $\mcl L_{\mcl R}$ did not converge.

From the results in Tab. \ref{tab:ablation_total} we conclude that without the frequency loss term, LIDS can achieve better fidelity, whereas it is less robust against the attacks that disturb the container image's high-frequency signal, such as JPEG compression, Low-pass filtering, and Gaussian blurring.
Surprisingly, this model could still resist color jitter attack and resizing and cropping well.
We believe that it was because such attacks had less impact on the high-frequency signal in the container image, and therefore, the model showed better robustness.

\paragraph{Clean loss.}
\begin{figure}[t!]
    \centering
    \includegraphics[width=.49\textwidth]{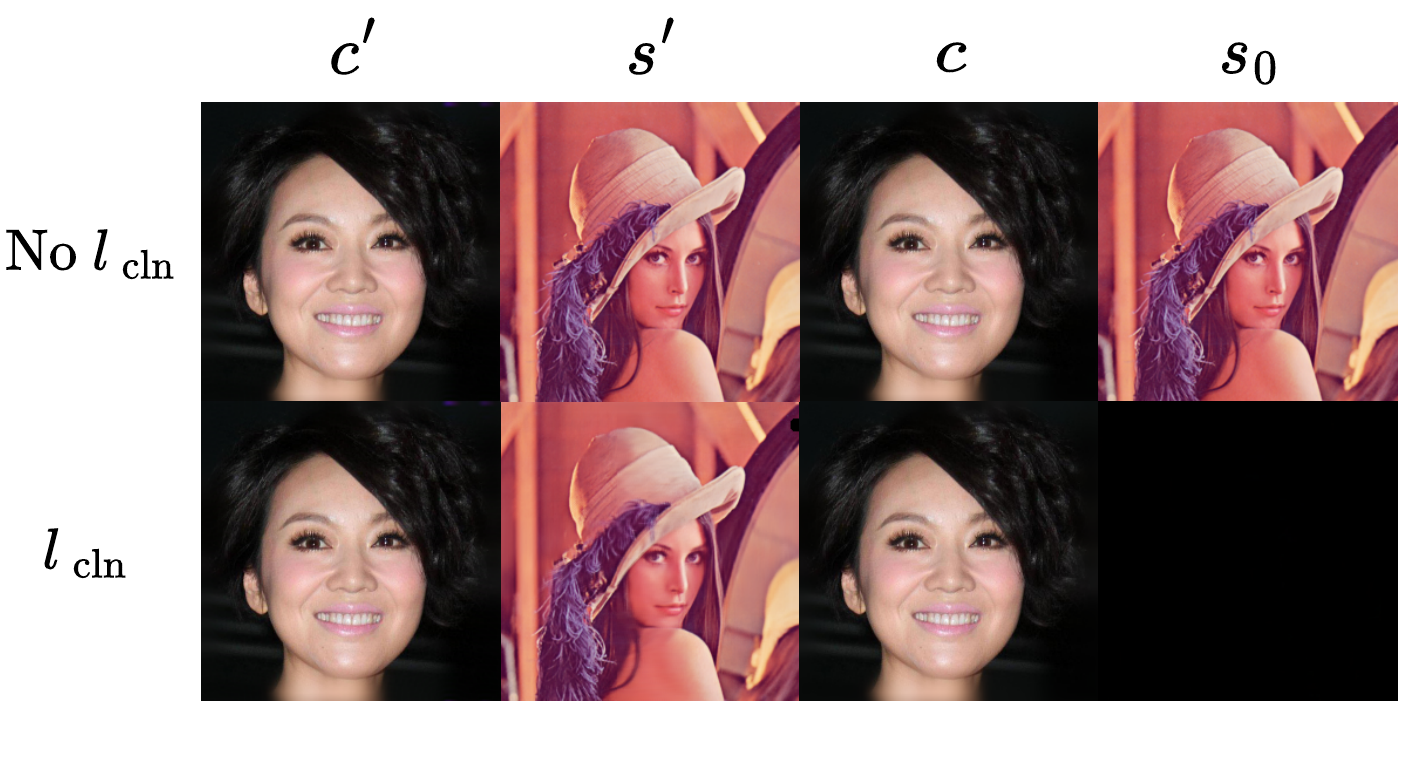}
    \caption{
        The Impact of Removing the Clean Loss.
        If the clean loss term is removed, the retrieval network will overfit to the secret image.
        This leads to that the retrieval network will output the secret image regardless of what it receives.
    }
    \label{fig:ablation_cln_loss}
\end{figure}
It is expected that if we remove the clean loss term from the loss function, the retrieval network will overfit to the secret image, resulting in a very low retrieval loss $l_{\text{ret}}$.
We thus train an LIDS model without the clean loss to exemplify our prediction.
We then feed arbitrary clean images to the model's retrieval network, and it retrieves the recovered secret image perfectly as shown in Fig. \ref{fig:ablation_cln_loss}, proving the necessity of the clean loss.

To further examine whether the clean loss is capable of enabling the model to distinguish container images from the clean images, we feed arbitrary clean images and their distorted counterparts to the retrieval network of a standard LIDS model.
This is due to the fact that the retrieval network is trained to retrieve the recovered  secret image from the attacked container images, and we are concerned that it may have memorized the attack patterns.
As depicted in Fig. \ref{fig:ablation_cln_loss}, the retrieval network is able to perfectly tell whether the given image has been watermarked or not.
As a result, the retrieval network can precisely determine whether or not an image is a container.





\section{Conclusion}
\label{sec:con}

In this paper, we introduced LIDS, a novel image watermarking method based on deep learning that embeds a secret image into the low-frequency components of cover images. Our experiments showed that LIDS has high fidelity, robustness, and specificity. The container images generated by LIDS had better visual quality compared to those produced by other methods, and the embedded secret image of LIDS survived intensive attacks, while the others failed. Moreover, LIDS could easily distinguish clean images from watermarked images, and it outperformed other state-of-the-art methods. LIDS can also take an arbitrary image as the watermark.
To the best of our knowledge, we are the first to use deep learning to embed a watermark into the low-frequency components of a protectee image while manipulating the frequency distribution of the extracted features. However, LIDS is not as effective when facing attacks that drastically change the geometrical information of the watermarked images, such as resizing and cropping. Therefore, it is necessary to develop methods that can withstand such attacks to further improve the watermark's robustness.

\clearpage

\bibliographystyle{IEEEtran}
\bibliography{egbib}

\clearpage

\begin{IEEEbiography}[{\includegraphics[width=1.2in,height=1.4in,clip,keepaspectratio]{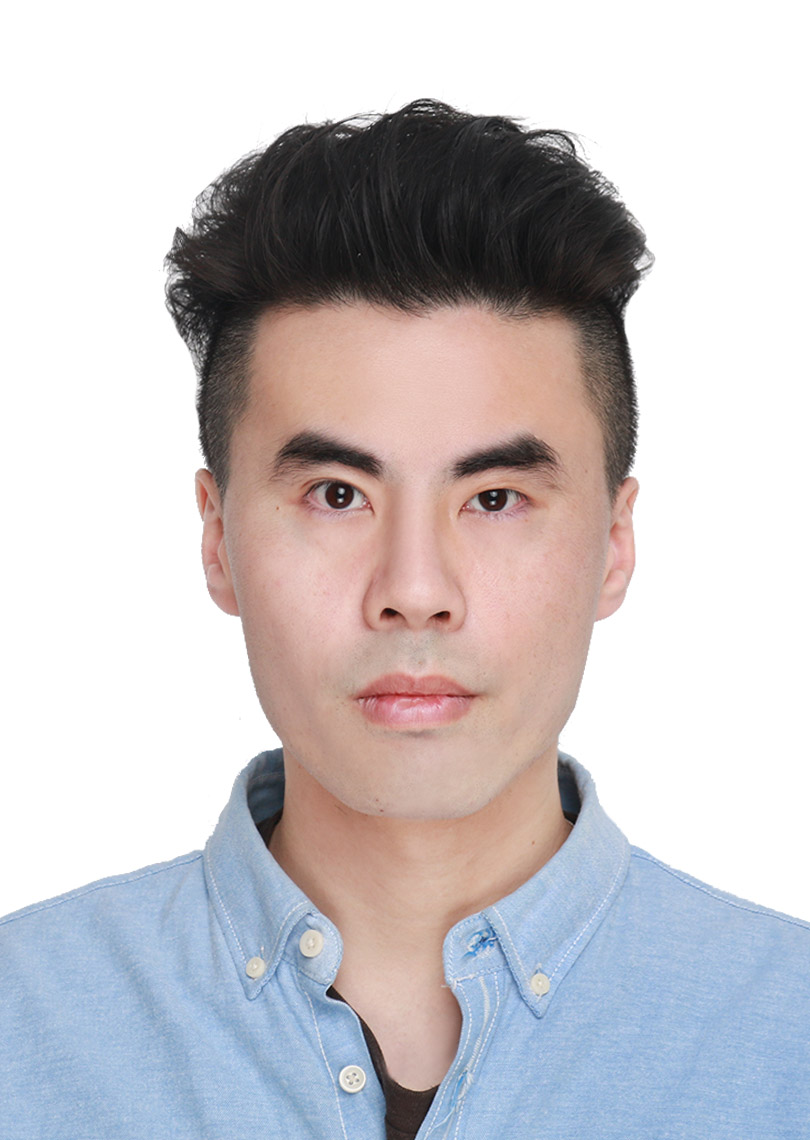}}]{Huajie Chen}
is currently a Ph.D. student at the School of Computer Science, University of Technology Sydney, Australia. He received his B.Sc. degree from Sun Yat-sen University, China, and M.Sc. degree from Universität Tübingen, Germany. His research interests include security and privacy in deep learning, cryptography, bioinformatics, etc.

\end{IEEEbiography}

\vspace{-10pt} 

\begin{IEEEbiography}[{\includegraphics[width=1in,height=1.25in,clip,keepaspectratio]{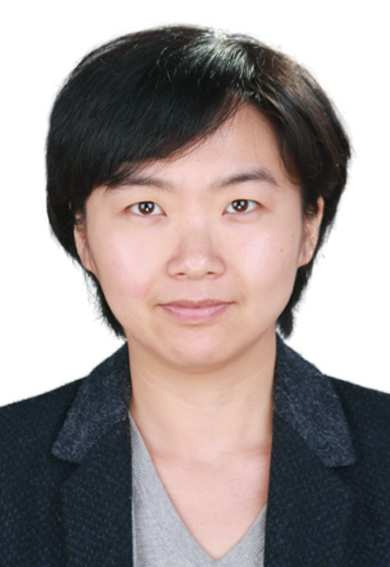}}]{Tianqing Zhu}
holds BEng and MEng degrees from Wuhan University, Wuhan, China and a PhD in Computer Science from Deakin University, Australia (2014). She is currently an Associate Professor with the School of Computer Science at the University of Technology Sydney, Australia. Prior to that, she was a Lecturer with the School of Information Technology, Deakin University, from 2014 to 2018. Her research interests include privacy-preserving, data mining, and network security.

\end{IEEEbiography}

\vspace{-10pt} 

\begin{IEEEbiography}[{\includegraphics[width=1in,height=1.25in,clip,keepaspectratio]{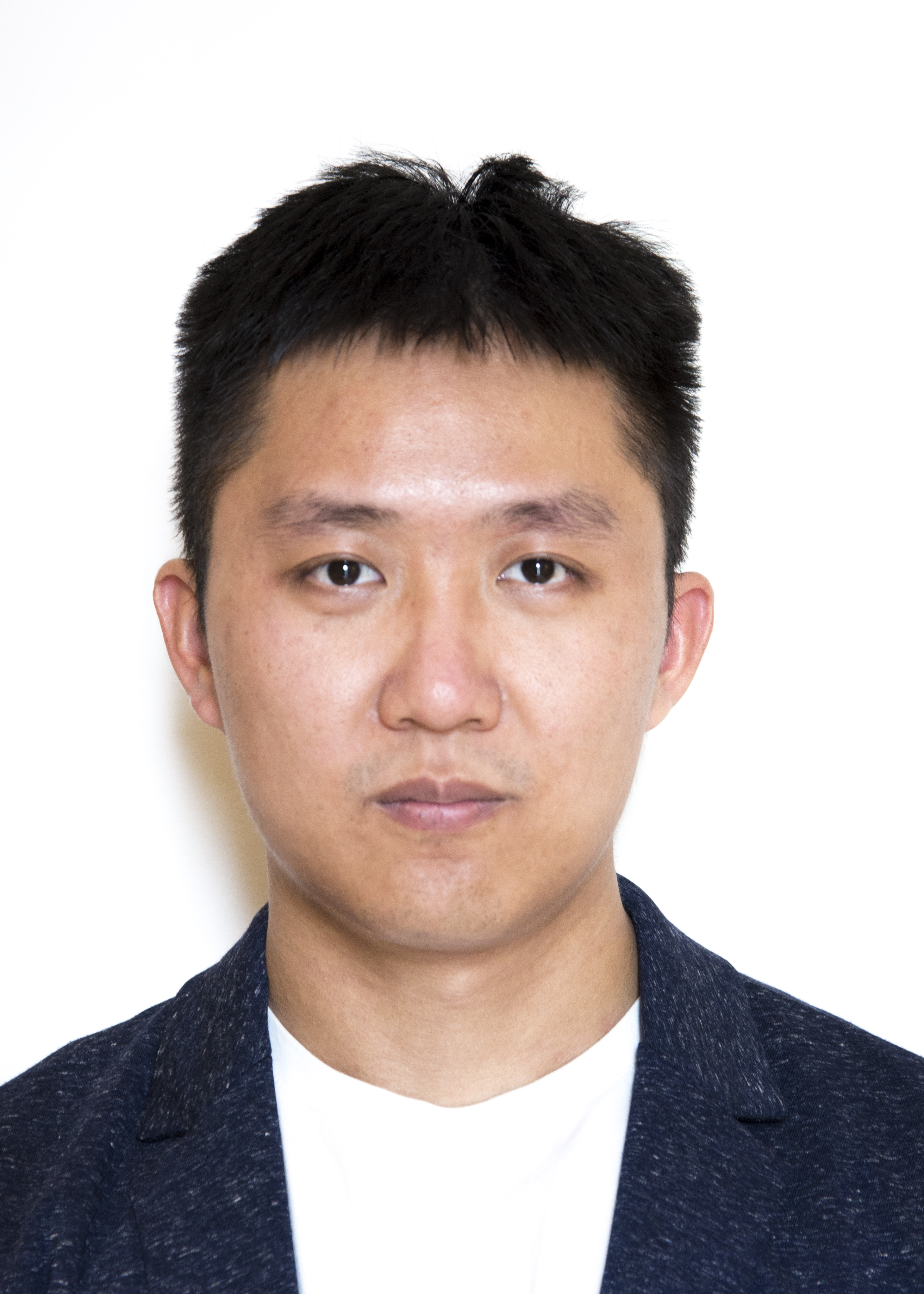}}]{Yuan Zhao}
received the B.Sc. degree in Internet of Things Engineering from The Northwest University, China, and the M.Sc. degree in Information Technology from The University of Sydney, Australia. He is currently pursuing his PhD degree at the School of Computer Science, University of Technology Sydney, Australia. His research interests focus on privacy and security in computer vision.

\end{IEEEbiography}

\vspace{-10pt}

\begin{IEEEbiography}[{\includegraphics[width=1in,height=1.25in,clip,keepaspectratio]{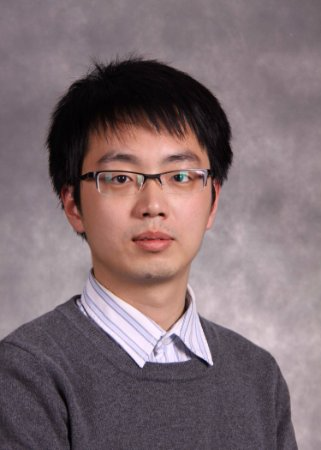}}]{Bo Liu}
received the BEng degree from the Department of Computer Science and Technology, Nanjing University of Posts and Telecommunications, Nanjing, China, in 2004. He then received the MEng. and PhD. Degrees from the Department of Electronic Engineering, Shanghai Jiao Tong University, Shanghai, China, in 2007 and 2010, respectively. He is currently a Senior Lecturer with the University of Technology Sydney, Australia. His research interests include cybersecurity and privacy, location privacy and image privacy, privacy protection and machine learning.

\end{IEEEbiography}

\vspace{-10pt} 

\begin{IEEEbiography}[{\includegraphics[width=1in,height=1.25in,clip,keepaspectratio]{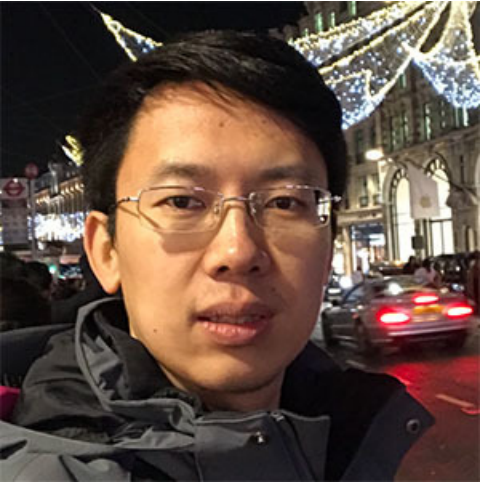}}]{Xin Yu}
received his B.S. degree in Electronic Engineering from University of Electronic Science and Technology of China, Chengdu, China, in 2009, and received his
Ph.D. degree in the Department of Electronic Engineering, Tsinghua University, Beijing, China, in 2015.
He also received a Ph.D. degree in the College of Engineering and Computer Science, Australian National University, Canberra, Australia, in 2019.
He is currently a lecturer in University of Technology Sydney.
His interests include computer vision and image processing.

\end{IEEEbiography}

\vspace{-10pt} 

\begin{IEEEbiography}[{\includegraphics[width=1in,height=1.25in,clip,keepaspectratio]{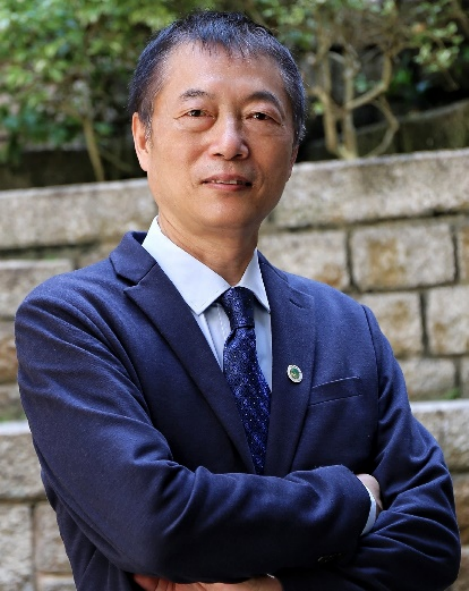}}]{Wanlei Zhou}
(Senior member, IEEE) is currently the Vice Rector (Academic Affairs) and Dean of Institute of Data Science, City University of Macau, Macao SAR, China. He received the PhD degree in Computer Science and Engineering from The Australian National University, Canberra, Australia, in 1991. He also received a DSc degree (a higher Doctorate degree) from Deakin University in 2002. Before joining City University of Macau, Professor Zhou held various positions including the Head of School of Computer Science in University of Technology Sydney, Australia, the Alfred Deakin Professor, Chair of Information Technology, Associate Dean, and Head of School of Information Technology in Deakin University, Australia. His main research interests include security, privacy, and distributed computing. He has published more than 400 papers in refereed international journals and refereed international conference proceedings, including many articles in IEEE transactions and journals.

\end{IEEEbiography}

\vfill

\end{document}